\pdfoutput=1 
\documentclass[preprint,aps,onecolumn,superscriptaddress,amsfont,graphicx,nofootinbib,preprintnumbers]{revtex4}
\usepackage{amsmath,amssymb,amsfonts}
\usepackage{bm}
\usepackage{soul}
\usepackage{color}
\usepackage{hyperref}
\usepackage{graphicx}
\usepackage{braket}
\usepackage{cancel}
\usepackage[english]{babel}
\usepackage[utf8]{inputenc}
\usepackage[T1]{fontenc}
\usepackage{float}
\usepackage{subfigure}
\usepackage{comment}
\DeclareMathSymbol{\shortminus}{\mathbin}{AMSa}{"39}

\definecolor{nicered}{rgb}{0.7,0.1,0.1}
\definecolor{nicegreen}{rgb}{0.0,0.4,0.0}
\hypersetup{colorlinks,citecolor=nicegreen,linkcolor=nicered,urlcolor=nicered}

\allowdisplaybreaks[1]
\newcommand{\as}{\alpha_s}
\newcommand{\aso}{\bar\alpha_{s}}
\newcommand{\asb}{\alpha_{s,b}}
\newcommand{\asbo}{\bar\alpha_{s,b}}
\newcommand{\ep}{\epsilon}

\newcommand{\be}{\begin{equation}}
\newcommand{\ee}{\end{equation}}

\newcommand*{\Scale}[2][4]{\scalebox{#1}{$#2$}}

\DeclareMathOperator{\tr}{Tr}
\DeclareMathOperator{\order}{\mathcal{O}}
\DeclareMathOperator{\myRe}{Re}
\DeclareMathAlphabet\mathbfcal{OMS}{cmsy}{b}{n}

\definecolor{darkred}{rgb}{0.9,0,0}

\definecolor{darkgreen}{rgb}{0,0,0.9}

\definecolor{darkblue}{rgb}{0,0,0.9}

\definecolor{darkpink}{rgb}{0.8,0.0,0.8}

\definecolor{orange}{rgb}{0.9,0.5,0}

\begin{document}

\def\OX{Rudolf Peierls Centre for Theoretical Physics, University of Oxford,\\
Clarendon Laboratory, Parks
Road, Oxford OX1 3PU, UK}
\def\TUM{Physik Department,
James-Franck-Straße 1, Technische Universit\"at M\"unchen,\\
D–85748 Garching, Germany}
\def\ORG{Exzellenzcluster ORIGINS, Boltzmannstr. 2, D-85748 Garching, Germany}
\def\MSU{Department of Physics and Astronomy, Michigan State University,\\
East Lansing, Michigan 48824, USA}
\def\WDM{Wadham College, University of Oxford, Parks Road, Oxford OX1 3PN, UK}
\def\NEW{New College, University of Oxford, Holywell Street, Oxford OX1 3BN, UK}
\preprint{OUTP-22-09P, MSUHEP-22-023, TUM-HEP-1404/22}

\title{Three-loop helicity amplitudes for quark-gluon scattering in QCD}

\author{Fabrizio Caola}            
\email[Electronic address: ]{fabrizio.caola@physics.ox.ac.uk}
\affiliation{\OX} 
\affiliation{\WDM}

\author{Amlan Chakraborty}            
\email[Electronic address: ] {chakra69@msu.edu}
\affiliation{\MSU}

\author{Giulio Gambuti}            
\email[Electronic address: ]{giulio.gambuti@physics.ox.ac.uk}
\affiliation{\OX}
\affiliation{\NEW}

\author{Andreas von Manteuffel}            
\email[Electronic address: ]{vmante@msu.edu}
\affiliation{\MSU}

\author{Lorenzo Tancredi}            
\email[Electronic address: ]{lorenzo.tancredi@tum.de}
\affiliation{\TUM}

\begin{abstract}
We compute the three-loop helicity amplitudes for $q\bar{q}\to gg$ and its crossed partonic channels, in massless QCD.
Our analytical results provide a non-trivial check of the color quadrupole contribution to the infrared poles for external states in different
color representations.
At high energies, the $qg \to qg$ amplitude shows the predicted factorized form from Regge theory and confirms previous results for the gluon Regge trajectory extracted from $qq'\to qq'$ and $gg \to gg$ scattering.
\end{abstract}

\maketitle 
\tableofcontents

\section{Introduction}
\noindent
The computation of multiloop scattering amplitudes in Quantum Chromodynamics (QCD) plays a fundamental role for the Standard Model (SM) precision program carried out at particle colliders such as the Large Hadron Collider (LHC) at CERN. Suitably combined with real-radiation contributions, they provide a powerful tool to generate predictions for a variety of collider observables, allowing for precise comparisons with experimental data~\cite{Heinrich:2020ybq}. In fact,
matching the shrinking experimental errors with correspondingly precise theory predictions allows one to discover even subtle signals from possible physics scenarios beyond the SM.

In addition to their phenomenological significance, analytic computations of scattering amplitudes enable investigations of general properties of perturbative Quantum Field Theories (QFT), including comparative studies of QCD amplitudes with their supersymmetric counterparts.
The more loops, external legs, or particle masses one is considering for a scattering amplitude, the more challenging its computation becomes.
In recent years, significant progress has been achieved for the reduction of loop integrals to master integrals and their analytical evaluation, resulting in the calculation of previously inaccessible multiloop amplitudes.
 At two loops, various QCD amplitudes became available for 2 $\rightarrow$ 3 scattering processes involving mostly massless particles~\cite{Badger:2017jhb,Abreu:2017hqn,Abreu:2018aqd,Abreu:2018zmy,Abreu:2018jgq,Abreu:2019rpt,Abreu:2020cwb,Chicherin:2018yne,Chicherin:2019xeg,Chawdhry:2020for,DeLaurentis:2020qle,Chawdhry:2018awn,Abreu:2020xvt,Agarwal:2021grm,Badger:2021nhg,Abreu:2021fuk,Agarwal:2021vdh,Chawdhry:2021mkw,Badger:2021imn,Gehrmann:2015bfy,Papadopoulos:2015jft,Gehrmann:2018yef,Chicherin:2018mue,Chicherin:2020oor,Abreu:2021asb,Badger:2022ncb}, paving the way for the first Next-to-Next-to-Leading-Order (NNLO) studies at LHC~\cite{Chawdhry:2019bji,Czakon:2021mjy,Chawdhry:2021hkp}.
 At three loops, first QCD amplitudes were computed for 2 $\rightarrow$ 2 scattering processes~\cite{Caola:2020dfu,Caola:2021rqz,Caola:2021izf,Bargiela:2021wuy}.
At four loops, $2 \to 1$ form factors were obtained in full-color QCD~\cite{Lee:2021uqq,Lee:2022nhh,Chakraborty:2022yan}.
  
Analytical results for multiloop scattering amplitudes
can also provide non-trivial information about all-order results in QCD. An interesting case is the
so-called Regge limit~\cite{Kuraev:1977fs} of large collision energy, where universal factorization properties can be observed in QCD amplitudes.
The BFKL formalism~\cite{Kuraev:1976ge,Balitsky:1978ic} allows one to describe all-order structures in QCD through the exchange of so-called ``Reggeized gluons'', which resum leading contributions of the quark and gluon interactions at high energies.
With the recent determination of the three-loop Regge trajectory~\cite{Caola:2021izf,Falcioni:2021dgr}, the last missing ingredient for next-to-next-to-leading-logarithmic analysis became available.

This paper concludes our analytical calculation of all four-parton scattering amplitudes in three-loop QCD.
Previously, we presented the helicity amplitudes
for the process $q\bar{q}\to q'\bar{q}'$ and crossed channels  \cite{Caola:2021rqz}
and for the process $gg\to gg$ \cite{Caola:2021izf}.
In this work, we provide the helicity amplitudes for $q\bar{q}\to gg$ scattering  and crossed channels in full-color, massless QCD.
Our calculation checks the predicted quadrupole contribution 
to the infrared poles for a process with external legs in different color representation~\cite{Almelid:2015jia,Henn:2016jdu}.
By analyzing the high-energy limit of the $qg\to qg$ amplitude, we check the universality of the predicted factorization and the three-loop expression for the Regge trajectory~\cite{Caola:2021izf,Falcioni:2021dgr}.

The rest of this paper is organized as follows.
In section~\ref{Kinematics}, we set up our notation and describe the color and Lorentz decomposition of the scattering amplitude. In section~\ref{Helicity} we
discuss our computation of the bare helicity amplitudes employing the tensor decomposition provided in the previous section and analytical solutions for the master integrals~\cite{Henn:2020lye,Bargiela:2021wuy}.
In section~\ref{UVIR}, we describe the UV renormalization and give details for the subtraction of IR poles up to three loops. In section~\ref{Checks} we present our final results and enumerate the checks we have performed to verify their correctness. Finally, in section~\ref{HighEn} we discuss the high energy (Regge) limit of the $qg\to qg$ amplitudes. We draw our conclusions in section~\ref{Conc}. We reserve the appendices for lengthy formulas with explicit results for all the relevant anomalous dimensions (appendix~\ref{app:andim}) and for the impact factors and the gluon 
Regge trajectory (appendix~\ref{app:impact}).

\section{Color and Lorentz decomposition}
\label{Kinematics}
\noindent
We consider the quark-gluon scattering process 
\begin{equation} 
  { q}(p_1) \;+ \;\bar { q}(p_2)  \; +\;  { g}(p_3) \;+ \; {   g}(p_4) \; \longrightarrow \; 0, 
\label{allincoming}
\end{equation}
in massless QCD, where the momenta satisfy
\begin{equation}
p_1^2=p_2^2=p_3^2=p_4^2=0, \qquad
p_1^\mu + p_2^\mu  + p_3^\mu + p_4^\mu = 0.
\end{equation}
The kinematics of the process eq.~\eqref{allincoming} can be parametrized in
terms of the usual Mandelstam invariants
\begin{align}
s
 = (p_1 + p_2)^2, \qquad
t 
 = (p_1 + p_3)^2, \qquad
u
 = (p_2 + p_3)^2,
\end{align}
with $u = -t-s$.
We find it convenient to introduce the dimensionless variable
\begin{equation}\label{variables}
x=-t/s 
\end{equation}
to parametrize our results.

The primary physical scattering process considered in this paper
is
\begin{equation}
  \label{s_channel} 
     {q}(p_1) \;+ \;\bar { q}(p_2)  \; \longrightarrow \; { g}(p_3) \;+ \; {   g}(p_4)\,,
\end{equation} 
which can be obtained from the process \eqref{allincoming} by a crossing of external legs
with $p_{3,4}\rightarrow-p_{3,4}$.
For this process, the physical region of the phase space is given by
\begin{equation}
\label{eq:physregion}
s >0,\quad t,u < 0
\qquad \Rightarrow \quad 0 < x < 1\,.
\end{equation}
Results for other physical scattering processes will subsequently
be derived from the result for process \eqref{s_channel} by
considering further crossings.
The bare amplitude for process~\eqref{s_channel} can be decomposed in three different color structures $\mathcal{C}_i$,
\begin{equation} \label{eq:decomp_color}
\mathcal{A}_{i_1,i_2,a_3,a_4}  = 4 \pi \alpha_{s,b}  \sum_{i=1}^3 \mathcal{A}^{[i]}  {\mathcal{C}}_{i}\,.
\end{equation}
Here, $i_1$ and $i_2$ are the fundamental color indices of the external quarks with momenta $p_1$ and $p_2$, and $a_3$ and $a_4$ are the adjoint color indices of the external gluons with momenta $p_3$ and $p_4$, respectively.
Further, $\alpha_{s,b}$ is the bare strong coupling.
In eq.~\eqref{eq:decomp_color} we also introduced the notation $[i]$ to indicate a color component index of the amplitude.
The three color structures are
\begin{align}
{\mathcal{C}}_{1} = ({{T}^{a_3}}{{T}^{a_4}})_{i_2i_1}, \quad \quad
{\mathcal{C}}_{2} = ({{T}^{a_4}}{{T}^{a_3}})_{i_2i_1},  \quad \quad
{\mathcal{C}}_{3} = \delta^{a_3a_4}\,\delta_{i_2 i_1}
\,, \label{eq:colorbasis}
\end{align}
where we work in QCD with color group $SU(N_c)$ and $n_f$ massless quark flavors. The matrices
$(T^a)_{i_2i_1}$ are the generators of $SU(N_c)$ in
the fundamental representation.
We use $\tr[T^aT^b] = \frac12 \delta_{ab}$ and denote
the quadratic Casimir operators in the fundamental
and adjoint representation by $C_F$ and $C_A$, respectively.

The amplitude coefficients $\mathcal{A}^{[i]}$ can be decomposed further into Lorentz-covariant structures $\mathcal{T}_i$,
\begin{equation}\label{eq:decomp_tensors}
\mathcal{A}^{[i]}  = \sum_{j=1}^{4}  \mathcal{F}^{[i]}_j \;  \mathcal{T}_i \;,
\end{equation}
where the $ \mathcal{F}^{[i]}_j$ are scalar form factors. 
To regulate ultraviolet and infrared divergences, we employ dimensional regularization and use $d = 4 - 2 \epsilon$ for the number of space-time dimensions.
We denote the external gluon polarization vectors as $\ep(p_i) = \ep_i $ with the transversality condition for the external gluon momenta $\ep(p_i)\cdot p_i = 0 $ ($i =3,4$). To simplify the Lorentz decomposition, we also
fix the gauge of the external gluons such that $\ep_3\cdot p_2 = \ep_4\cdot p_1 = 0 $, which leads to the following gluon polarization sums 
\begin{align}
\label{Polsum}
&\sum_{pol}{\ep}^{\mu}_3 {\ep}^{\nu}_3 = -g^{\mu\nu} + \frac{{p}^{\mu}_3{p}^{\nu}_2 + {p}^{\nu}_3{p}^{\mu}_2}{p_2\cdot p_3},\,\nonumber\\
&\sum_{pol}{\ep}^{\mu}_4 {\ep}^{\nu}_4 = -g^{\mu\nu} + \frac{{p}^{\mu}_4{p}^{\nu}_1 + {p}^{\nu}_4{p}^{\mu}_1}{p_1\cdot p_4}\,.
\end{align}
Since we are ultimately interested in computing the helicity amplitudes for this process in the ’t Hooft--Veltman scheme (tHV) scheme, we use the Lorentz structures \cite{Peraro:2019cjj,Peraro:2020sfm, Caola:2020dfu}
\begin{alignat}{2}\label{lorentzstructs}
\mathcal{T}_1 &= \bar{u}(p_2){\cancel{\ep}}_{3}u(p_1) \,\ep_4\cdot p_2\,,  \quad \quad
&&\mathcal{T}_2 = \bar{u}(p_2){\cancel{\ep}}_{4}u(p_1) \,\ep_3\cdot p_1\,, \nonumber\\
\mathcal{T}_3 &= \bar{u}(p_2){\cancel{p}}_{3}u(p_1) \,\ep_3\cdot p_1\,\ep_4\cdot p_2\,, \quad \quad
&&\mathcal{T}_4 = \bar{u}(p_2){\cancel{p}}_{3}u(p_1) \,\ep_3\cdot \ep_4\,,
\end{alignat}
and introduce projection operators $\mathcal{P}_i$ which extract the form factors from the amplitude,

\begin{equation}
\mathcal{P}_{j}\cdot \mathcal{A}^{[i]} = \sum_{\text{pol}}\mathcal{P}_{j}\mathcal{A}^{[i]} = \mathcal{F}^{[i]}_j,\quad j = 1,\ldots,4\,. \label{eq:defproj}
\end{equation}
In eq.~\eqref{eq:defproj}, we introduced the short-hand notation $\mathcal{P}_{i}\cdot \mathcal{A}$ which implies a sum over the polarizations of the external particles.
By introducing the matrix 
$$M_{ij} =\mathcal{T}^\dagger_i \cdot \mathcal{T}_j,$$
the projectors can be compactly defined as
\begin{align}
\mathcal{P}_{i} = \sum_{j=1}^{4}(M^{-1})_{ij}\mathcal{T}^{+}_{j} \quad \implies    \label{eq:projectors} \quad
\mathcal{P}_i\cdot \mathcal{T}_j = \delta_{ij} \,,
\end{align}
where
\begin{align}
M^{-1} &= \frac{1}{2(d-3)s^2 t^3 u}
\left(
\begin{array}{cccc}
t^2u^2 & 0 & -t u^2 & 0 \\
0 & t^2 u^2 & t u^2 & 0 \\
-t u^2 & t u^2 & \,\,(d u^2-4st)\,\, & (s-t)s t \\
0 & 0 & (s-t)s t & s^2 t^2 \\
\end{array}
\right)\,.\label{eq:matrixprojqqbaa2}
\end{align}
We stress that in conventional dimensional regularization there is a fifth Lorentz structure which would need to be taken into account in eq.~\eqref{eq:decomp_tensors}.
In the tHV scheme we take internal momenta in $d = 4 - 2 \epsilon$ dimensions and keep external momenta and polarizations in four dimensions.
As explained in refs.~\cite{Peraro:2019cjj,Peraro:2020sfm},
this allows us to essentially ignore this fifth evanescent structure completely and work with just
the four structures \eqref{lorentzstructs}, which
are linearly independent in four space-time dimensions.
We also point out that the decompositions of eqs.~\eqref{eq:decomp_color} and \eqref{eq:decomp_tensors},
as well as the explicit form of the projectors \eqref{eq:projectors}, hold to any orders in perturbation theory.

\section{Helicity amplitudes}
\label{Helicity}
\noindent
From the form factors $\mathcal{F}_j$ one can construct amplitudes for definite helicities
of the external particles.
We denote the helicity of the incoming quark as $\lambda_q$; the helicity of the incoming anti-quark $\lambda_{\bar{q}}$ is then automatically fixed due to helicity conservation along the massless quark line.
We refer to the quark line helicity with the symbol $\lambda_{q\bar{q}} = \{\lambda_{q}\lambda_{\bar{q}} \}$ which can take two possible values: $\lambda_{q\bar{q}}  = L,R = \{-+\},\{+-\} $.
Further, we denote the helicities of the outgoing gluons as $\lambda_{3}$ and $\lambda_4$.
After exploiting parity, charge-conjugation and Bose symmetry relations~\cite{Caola:2020dfu}, one is left with only two independent helicity configurations.
However, we choose to compute the overcomplete set of four helicity configurations
\begin{equation} 
\{\lambda_{q\bar{q}}\lambda_{3}\lambda_{4}\}
= \{L--\}, \{L-+\}, \{L+-\}, \{L++\}
\label{eq:refhelicities}
\end{equation}
which allow us to perform a consistency check on our calculation. 
Results for right-handed quarks can subsequently be obtained by a parity transformation.
We write for the left-handed spinors
$
\overline{u_{L}}(p_2) = \langle 2|$, 
$u_L(p_1) = |1] $,
and for the polarization vector of the gluons
\begin{align}
\ep^{\mu}_{3,-}(p_3) &= \frac{\langle 2|\gamma^{\mu}|3]}{\sqrt{2}\langle 23 \rangle}\,, &
\ep^{\mu}_{3,+}(p_3) &= \frac{\langle 3|\gamma^{\mu}|2]}{\sqrt{2}[32]}\,,
\\
\ep^{\mu}_{4,-}(p_4) &= \frac{\langle 1|\gamma^{\mu}|4]}{\sqrt{2}\langle 14 \rangle}\,, &
\ep^{\mu}_{4,+}(p_4) &= \frac{\langle 4|\gamma^{\mu}|1]}{\sqrt{2}[41]}\,.
\end{align}
Inserting these equations into the Lorentz structures $\mathcal{T}_j$ \eqref{lorentzstructs}
gives the helicity amplitudes
\begin{align}
\label{helicitydef}
\mathcal{A}_{L--}  &= s_{L--} \sum_{i=1}^3 \mathcal{H}^{[i]}_1 \;{\mathcal{C}}_{i}\,, \quad
&\mathcal{A}_{L-+} &= s_{L-+}\sum_{i=1}^3 \mathcal{H}^{[i]}_2\;{\mathcal{C}}_{i}\,, \nonumber \\
\mathcal{A}_{L+-}  &= s_{L+-} \sum_{i=1}^3 \mathcal{H}^{[i]}_3\;{\mathcal{C}}_{i}\,, \quad
&\mathcal{A}_{L++}  &= s_{L++} \sum_{i=1}^3 \mathcal{H}^{[i]}_4\;{\mathcal{C}}_{i}\,, 
\end{align}
where the little group scaling is captured by the overall spinor factors
\begin{align}\label{eq:spinor_factors}
s_{L--} = \frac{2[34]^2}{\langle 1 3 \rangle [23]}\,, \quad
s_{L-+} = \frac{2 \langle 2 4 \rangle  [13]}{\langle 2 3 \rangle [24]}\, , \quad
s_{L+-} = \frac{2\langle 2 3 \rangle  [41]}{\langle 2 4 \rangle [32]}\, , \quad
s_{L++} = \frac{2 {\langle 3 4 \rangle}^2}{\langle 3 1 \rangle [23]}\,,
\end{align}
and we have defined the scalar helicity amplitudes
\begin{align}
\label{FFtoH}
\mathcal{H}^{[i]}_1 &= \frac{t}{2}\left(\mathcal{F}^{[i]}_2 - \frac{t}{2}\mathcal{F}^{[i]}_3 + \mathcal{F}^{[i]}_4 \right),  &\mathcal{H}^{[i]}_2 &=  \frac{t}{2}\left(\frac{s}{2}\mathcal{F}^{[i]}_3 + \mathcal{F}^{[i]}_4 \right),& \nonumber\\
\mathcal{H}^{[i]}_3 &= \frac{st}{2u}\left(\mathcal{F}^{[i]}_2 - \mathcal{F}^{[i]}_1 - \frac{t}{2}\mathcal{F}^{[i]}_3 - \frac{t}{s}\mathcal{F}^{[i]}_4 \right), & 
\mathcal{H}^{[i]}_4 &= \frac{t}{2}\left(\mathcal{F}^{[i]}_1 + \frac{t}{2}\mathcal{F}^{[i]}_3 - \mathcal{F}^{[i]}_4 \right).&
\end{align}
The amplitudes for right-handed quarks are related to those for left-handed quarks by 
\begin{equation}
\mathcal{A}_{R,\lambda_3,\lambda_4} = ( \mathcal{A}_{L,-\lambda_3,-\lambda_4} )|_{\langle ij \rangle \leftrightarrow [ji]}\,.
\end{equation}
By exchanging the two outgoing gluons, we find that Bose symmetry implies the relations
\begin{alignat}{6}
&\mathcal{H}_2^{[1]}(x)&=&+\mathcal{H}_3^{[2]}(1-x), \qquad
&&\mathcal{H}_2^{[2]}(x)&=&+\mathcal{H}_3^{[1]}(1-x), \qquad
&&\mathcal{H}_2^{[3]}(x)&=&+\mathcal{H}_3^{[3]}(1-x),  \nonumber\\
&\mathcal{H}_{1,4}^{[1]}(x)&=& -\mathcal{H}_{1,4}^{[2]}(1-x), \qquad
&&\mathcal{H}_{1,4}^{[2]}(x)&=& -\mathcal{H}_{1,4}^{[1]}(1-x), \qquad
&&\mathcal{H}_{1,4}^{[3]}(x)&=& -\mathcal{H}_{1,4}^{[3]}(1-x).
\end{alignat}
We also note that
\begin{equation}
    \mathcal{H}_1^{[i]}(x) = - \mathcal{H}_4^{[i]}(x) \, . 
\end{equation}
These identities will serve as an important check of our calculations.

We expand the helicity amplitudes in
$\asbo \equiv {\asb}/({4\pi})$,
\begin{align}\label{full_amplitude}
\mathcal{H}^{[i]}_\lambda &= \sum_{\ell=0}^3 \mathcal{H}^{[i],(\ell)}_\lambda \left( \asbo S_\epsilon \right)^\ell 
+ \order\left( \asbo^4 \right)
\end{align} 
for $\lambda=1,\ldots,4$, where $S_\ep = (4 \pi)^\ep e^{- \ep \gamma_E}$.
The normalization factor $S_\epsilon$ absorbs constants in the bare amplitude and matches the usual $\overline{\text{MS}}$ conventions in the renormalization of the strong coupling performed below.
In the expansion of the amplitude, $\mathcal{H}^{[i],(3)}_\lambda$ is the three loop contribution, which we compute here for the first time.
We have also recomputed the tree-level, one-loop and two-loop contributions using the form factor decomposition defined in eq.~\eqref{lorentzstructs}.

\begin{figure}
\centering
\subfigure{\includegraphics[width=.25 \textwidth]{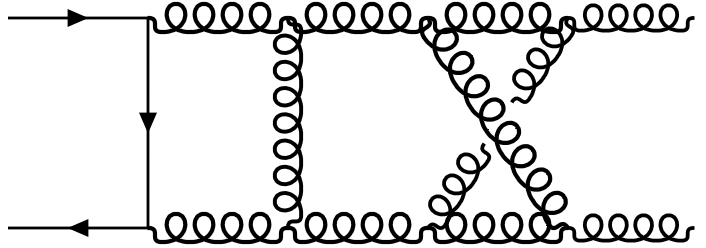}}\quad\quad\;\;\;
\subfigure{\includegraphics[width=.25 \textwidth]{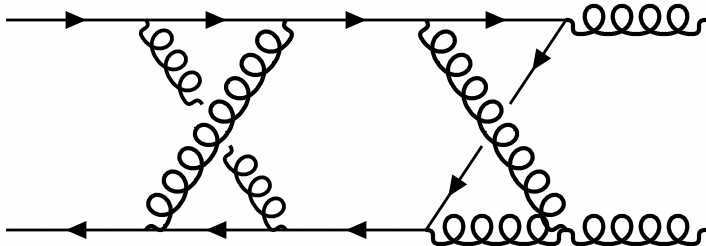}}
\caption{Sample three loop diagrams
contributing to the process $q\bar q \rightarrow gg $.} \label{diagrams}
\end{figure}

We employ \texttt{Qgraf}~\cite{Nogueira:1991ex} to produce Feynman diagrams and find 3 diagrams at tree level, 30 diagrams at one loop, 595 diagrams at two loops and 14971 at three loops. We give a few representative samples of the three-loop diagrams contributing to the process in figure~\ref{diagrams}.

\begin{figure}
\centering
\,\,\subfigure{\includegraphics[width=.249 \textwidth]{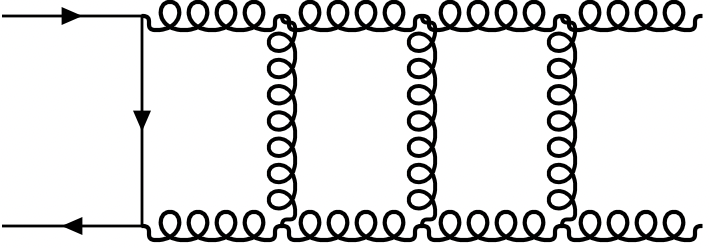}}\quad\quad\quad
\;\subfigure{\includegraphics[width=.252 \textwidth]{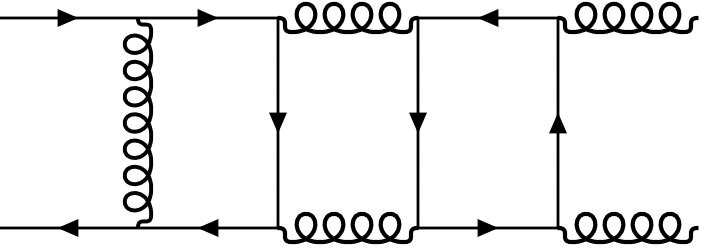}}
\end{figure}
We use \texttt{Form} \cite{Vermaseren:2000nd} to apply the Lorentz projectors of eq.~(\ref{lorentzstructs}) to the diagrams and to perform the Dirac and color algebra.
In this way, we obtain the form factors as linear combinations of a large number $(\sim10^7)$ of scalar Feynman integrals with rational coefficients. 
We parametrize the corresponding $\ell$-loop Feynman integrals according to
\begin{equation}\label{integrals}
\mathcal{I}^\text{top}_{n_1,n_2,...,n_N} = \mu_0^{2\ell \epsilon} e^{\ell \epsilon \gamma_E}  \int \prod_{j=1}^\ell \left( \frac{\mathrm{d}^d k_j}{i \pi^{\frac{d}{2}}} \right) \frac{1}{D_1^{n_1}D_2^{n_2} \dots D_N^{n_N}} \; ,
\end{equation}
where $\gamma_E \approx 0.5772$  is Euler's constant, $\mu_0$ is the scale of dimensional regularization,
and the denominators $D_j$ are inverse propagators for the respective integral family ``\text{top}''.
More details on the integral families can be found in ref.~\cite{Caola:2021rqz}. Using \texttt{Reduze 2}~\cite{Studerus:2009ye,vonManteuffel:2012np} and \texttt{Finred}, an in-house implementation of the Laporta algorithm~\cite{Laporta:2001dd} based on finite field arithmetic~\cite{vonManteuffel:2014ixa,
  vonManteuffel:2016xki,Peraro:2016wsq,Peraro:2019svx} and syzygy algorithms~\cite{Gluza:2010ws,Schabinger:2011dz,Ita:2015tya,Larsen:2015ped,Boehm:2017wjc,Agarwal:2019rag}, we reduced these integrals to a linear combination of 486 master integrals.
  Upon insertion of the recently computed solutions for the master integrals~\cite{Henn:2020lye,Bargiela:2021wuy} we arrive at an analytical result for the helicity amplitudes in terms of harmonic polylogarithms.

\section{UV and IR subtractions} \label{UVIR}
\begin{figure}
\centering
\subfigure[]{\includegraphics[width=0.23\textwidth]{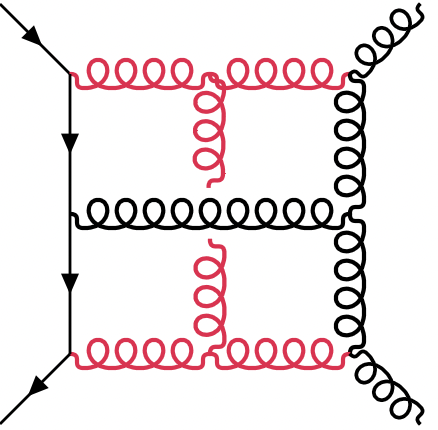}}\label{fig:2a}\quad\quad\quad\quad
\subfigure[]{\includegraphics[width=0.24\textwidth]{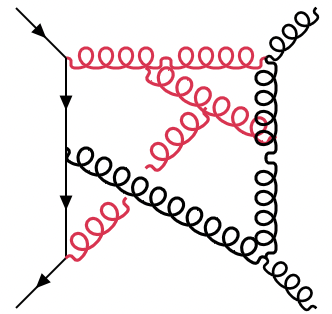}}\label{fig:2b}
\caption{Sample diagrams with quadrupole soft divergences, reinterpreted as tree-level diagrams (black lines) plus virtual gluons (red lines). Diagrams (a) and (b) involve colour correlations between four and three external partons and contribute to the first and second line of eq.~\eqref{eq:quadrupole}, respectively.} \label{diagrams_red}
\end{figure}

\noindent
The bare helicity amplitudes~\eqref{full_amplitude} contain UV and IR  divergences, which appear as poles in the Laurent expansion in $\ep$.
The $\overline{\text{MS}}$ renormalized strong $\as(\mu)$ is defined through
\begin{align}
\label{bare_to_phys}
\asbo \: \mu_0^{2\epsilon} \: S_\epsilon &= \aso \: \mu^{2\epsilon} Z[\aso] \; ,
\end{align}
where $\aso=\alpha_s(\mu)/(4\pi)$,  $\mu$ is the renormalization scale and
\begin{align}
Z[\aso]  =  1 -  \aso \frac{ \beta_0 }{\epsilon } + \aso^2 \left( \frac{\beta_0^2}{\epsilon^2} - \frac{\beta_1 }{2 \epsilon} \right)  -\aso^3  \left( \frac{\beta_0^3}{\epsilon^3} - \frac{ 7}{6} \frac{\beta_0 \beta_1}{\epsilon^2}+ \frac{\beta_2}{3 \epsilon} \right) + \mathcal{O}(\aso^4).
\end{align}
The $\beta$-function coefficients are defined in the standard way through
\begin{equation}
\frac{d \aso }{d \log \mu}  = \beta(\aso) - 2\epsilon \aso \; , \quad
\beta(\aso) = -2 \aso \sum\limits_{\ell \geq 0} \beta_\ell \aso^{\ell+1}.
\end{equation}
We also recall the values of the standard quadratic Casimir constants for a $SU(N_c)$ gauge group: 
\begin{equation}
    C_A = N_c, \qquad  C_F = \frac{N_c^2-1}{2N_c} \, .
\end{equation}
With this, up to third order of the perturbative expansion, we have
\begin{align}
\beta_0 &= \frac{11}{3} C_A - \frac{2}{3}\: n_f \; , \nonumber\\
\beta_1 &= \left( \frac{34}{3}  C_A^2-\frac{10}{3} C_A \:n_f \right)-2 \:C_F\: n_f \; ,\nonumber\\ 
\beta_2 &= -\frac{1415}{54}{C_A^2 \:n_f}+\frac{2857}{54}{ C_A^3}-\frac{205}{18}C_A\: C_F\:
   n_f +\frac{79}{54}C_A \:n_f^2+C_F^2 \:n_f+\frac{11}{9}C_F \:n_f^2 \;.
\end{align}
In the following, we use boldface symbols to denote vectors in colour space,
that is, we define 
\begin{equation}
\mathbfcal{H}= \big(\mathcal{H}^{[1]},\mathcal{H}^{[2]},\mathcal{H}^{[3]}\big)^T
\end{equation}
for the decomposition of the amplitude with respect to the basis $\mathcal{C}_i$.
Using the expansion of~\eqref{full_amplitude}, we collect the $\aso$ coefficients of the UV finite, but IR divergent, amplitudes as 
\begin{align}\label{hel_ampls_ren}
\mathbfcal{H}_{\lambda,\text{ren}}^{(0)} &=  \mathbfcal{H}_\lambda^{(0)} ,& \nonumber\\
\mathbfcal{H}_{\lambda,\: \text{ren}}^{(1)} &= \mathbfcal{H}_\lambda^{(1)}- \frac{\beta_0 }{\epsilon}   \mathbfcal{H}_\lambda^{(0)}, & \nonumber\\ 
\mathbfcal{H}_{\lambda,\: \text{ren}}^{(2)} &=   \mathbfcal{H}_\lambda^{(2)}  - \frac{2 \beta_0 }{\epsilon} \mathbfcal{H}_\lambda^{(1)}  + \frac{ \left(2 \beta_0^2- \beta_1\epsilon\right)}{2 \epsilon^2} \mathbfcal{H}_\lambda^{(0)}, & \nonumber\\ 
\mathbfcal{H}_{\lambda,\: \text{ren}}^{(3)} &= \mathbfcal{H}_\lambda^{(3)} -\frac{3 \beta_0}{\epsilon} \mathbfcal{H}_\lambda^{(2)} +\frac{ \left(3 \beta_0^2-\beta_1 \epsilon\right)}{\epsilon^2}  \mathbfcal{H}_\lambda^{(1)} +\frac{ \left(7 \beta_1 \beta_0
   \epsilon -6 \beta_0^3-2 \beta_2 \epsilon^2 \right)}{6 \epsilon^3} \mathbfcal{H}_\lambda^{(0)},
\end{align}
so that the renormalized helicity amplitudes can be written as 
$$\mathbfcal{H}_{\lambda,\text{ren}}=\sum_{\ell \geq 0}\aso^\ell \mathbfcal{H}^{(\ell)}_{\lambda,\text{ren}}\,.$$
The IR singularity structure of QCD amplitudes has been studied at two loops in ref.~\cite{Catani:1998bh} and was extended up to three loops in refs.~\cite{Sterman:2002qn,Aybat:2006wq,Aybat:2006mz,Becher:2009cu,Becher:2009qa,Dixon:2009gx,Gardi:2009qi,Gardi:2009zv,Almelid:2015jia}.
The IR divergences can be subtracted from our renormalized amplitudes multiplicatively:
\begin{equation}\label{eq:IR_factorization}
\mathbfcal{H}_{\lambda,\:\text{ren}}= \mathbfcal{Z} \; \mathbfcal{H}_{\lambda,\:\text{fin}}.
\end{equation}
Here $\mathbfcal{Z}$ is a color matrix acting on the space spanned by the $\mathcal C_i$ basis vectors \eqref{eq:colorbasis} and $\mathbfcal{H}_{\lambda,\:\text{fin}}$ are finite remainders, also called hard scattering functions.
The matrix $\mathbfcal{Z}$ can be written as 
\begin{equation}\label{exponentiation_B}
\mathbfcal{Z} = \mathbb{P} \exp \left[
  \int_\mu^\infty \frac{\mathrm{d} \mu'}{\mu'}
  \mathbf{\Gamma}(\{p\},\mu')\right], 
\end{equation}
where $\mathbb{P}$ denotes the path-ordering
of color operators~\cite{Becher:2009qa} in increasing values of $\mu'$ from left to
right. It can be omitted up to three loops,
since to this order $[\mathbf\Gamma(\mu),\mathbf\Gamma(\mu')] = 0$.
The color-space correlation structure at three-loops
allows one to decompose the soft anomalous dimension operator $\mathbf{\Gamma}$ into so-called \emph{dipole} ($\mathbf{\Gamma}_\text{dipole}$)
and \emph{quadrupole} ($\mathbf{\Delta}_4$) contributions according to 
\begin{equation}\label{anomalous_operator_B}
\mathbf{\Gamma} =  \mathbf{\Gamma}_\text{dipole}  + \mathbf{\Delta}_4 \, .
\end{equation}
The dipole term $\mathbf{\Gamma}_\text{dipole}$ can be written as
\begin{align}\label{dipole_B}
\mathbf{\Gamma}_{\text{dipole}}(\{p\},\mu)  = & \sum_{1\leq i < j \leq 4} \mathbf{T}^a_i \; \mathbf{T}^a_j  \; \gamma^\text{K}(\aso) \; \log\left(\frac{\mu^2}{-s_{ij}-i\delta}\right) + \sum_{i=1}^4 \;  \gamma^i(\aso)\; , & 
\end{align}
where $\gamma^\text{K}(\aso)$ is the \emph{cusp anomalous
dimension} \cite{Korchemsky:1987wg,Moch:2004pa,Vogt:2004mw,Bruser:2019auj,Henn:2019swt,vonManteuffel:2020vjv} and $\gamma^i$ the quark (gluon) \emph{collinear anomalous dimension} \cite{Ravindran:2004mb,Moch:2005id,Moch:2005tm,Agarwal:2021zft} of the $i$-th
external particle, which are given in our notation in appendix~\ref{app:andim}.
Further, $\mathbf{T}^a_i$ represents the color generator of the $i$-th parton in the scattering amplitude,
\begin{alignat}{2}\label{convention}
(\mathbf{T}^a_i)_{\alpha\beta} &=  t^a_{\alpha \beta} \; &&\text{ for a final(initial)-state quark (anti-quark)}, \nonumber\\
(\mathbf{T}^a_i)_{\alpha\beta} &=  -t^a_{\beta\alpha} \; &&\text{ for a final(initial)-state anti-quark (quark)}, \nonumber\\
(\mathbf{T}^a_i)_{bc} &=  -if^{abc} \;
&&\text{ for a gluon.}
\end{alignat}
The quadrupole term $\mathbf{\Delta}_4$ contributes for the first time at three loops. It can be written in the kinematical region \eqref{eq:physregion} as \cite{Almelid:2015jia,Henn:2016jdu,Caola:2021rqz,Caola:2021izf}
\begin{align} \label{eq:quadrupole}
\mathbf{\Delta}^{(3)}_4 &= 
128f_{abe}f_{cde}\left[ \mathbf{T}^a_1\mathbf{T}^c_2\mathbf{T}^b_3\mathbf{T}^d_4\,D_1(x) - \mathbf{T}^a_4\mathbf{T}^b_1\mathbf{T}^c_2\mathbf{T}^d_3\,D_2(x) \right]  \notag
\\
&- 16f_{abe}f_{cde}C\sum_{i=1}^4 \sum_{\substack{1\leq j < k \leq4 \\ j,k\neq i}}  \left\{ \mathbf{T}^a_i,\mathbf{T}^d_i \right\}\mathbf{T}^b_j\mathbf{T}^c_k,
\end{align}
where $C$ = $ \zeta_5 + 2 \zeta_2 \zeta_3$ and $D_1(x)$, $D_2(x)$ are linear combinations of harmonic polylogarithms as~\cite{Almelid:2015jia,Henn:2020lye,Caola:2021rqz,Caola:2021izf}. They read
\begin{align}
D_1 &= -2 \textit{G}_{1,4}-\textit{G}_{2,3}-\textit{G}_{3,2}+2 \textit{G}_{1,1,3}+2 \textit{G}_{1,2,2}-2 \textit{G}_{1,3,0}-\textit{G}_{2,2,0}-\textit{G}_{3,1,0} +2 \textit{G}_{1,1,2,0}\nonumber\\
&\quad -2 \textit{G}_{1,2,0,0}+ 2 \textit{G}_{1,2,1,0}+4 \textit{G}_{1,0,0,0,0}-2 \textit{G}_{1,1,0,0,0}+\frac{\zeta_5}{2}  - 5 \zeta_2 \zeta_3 + \zeta_2[5 \textit{G}_{3}+5 \textit{G}_{2,0}+2 \textit{G}_{1,0,0}\nonumber\\
 &\quad -6 (\textit{G}_{1,2}+\textit{G}_{1,1,0})]+ \zeta_3 (\textit{G}_{2}+2 \textit{G}_{1,0}-2 \textit{G}_{1,1}) 
 - i \pi  [-\zeta_3 \textit{G}_{0}+\textit{G}_{2,2}+\textit{G}_{3,0} +\textit{G}_{3,1}+ \textit{G}_{2,0,0}\nonumber\\
&\quad +2 (\textit{G}_{1,3}-\textit{G}_{1,1,2}-\textit{G}_{1,2,1} -\textit{G}_{1,0,0,0})] + i \pi \zeta_2 (-\textit{G}_{2}+2 (\textit{G}_{1,1}+\textit{G}_{1,0}))- 11i\pi \zeta_4 \, ,\label{D1}\\[10pt]
D_2 &=  2 \textit{G}_{2,3}+2 \textit{G}_{3,2}-\textit{G}_{1,1,3}-\textit{G}_{1,2,2}-2 \textit{G}_{2,1,2}+2 \textit{G}_{2,2,0}-2 \textit{G}_{2,2,1}  +2 \textit{G}_{3,1,0}-2 \textit{G}_{3,1,1}-\textit{G}_{1,1,2,0}\nonumber\\
&\quad - \textit{G}_{1,2,1,0}-2 \textit{G}_{2,1,1,0}+4 \textit{G}_{2,1,1,1}-\zeta_5 +4 \zeta_2 \zeta_3 + \zeta_3 \textit{G}_{1,1} +\zeta_2 [-6 \textit{G}_{3}-6 \textit{G}_{2,0}+2 \textit{G}_{2,1} \nonumber\\
&\quad +5 (\textit{G}_{1,2}+\textit{G}_{1,1,0})] + i \pi  (\zeta_3 \textit{G}_{1}+2 \textit{G}_{3,0}-\textit{G}_{1,1,2}-\textit{G}_{1,2,0}-\textit{G}_{1,2,1}+2 \textit{G}_{2,0,0}-2 \textit{G}_{2,1,0} \nonumber\\
&\quad+2 \textit{G}_{2,1,1}-\textit{G}_{1,1,0,0})+i \pi\zeta_2  (4 \textit{G}_{2}-\textit{G}_{1,1}) \, .  \label{D2}
\end{align}
Here the argument $x$ has been suppressed, and for the HPLs we used a compact notation similar to~\cite{Remiddi:1999ew,Maitre:2005uu}: 
\begin{align}
G_{a_1,\dots,a_n,\footnotesize\underbrace{  0,\dots,0}_{n_0}} = G(\underbrace{0,\dots,0}_{|a_1|-1},\text{sgn}(a_1),\dots,\underbrace{0,\dots,0}_{|a_n|-1},\text{sgn}(a_n),\underbrace{0,\dots,0}_{n_0};x). \nonumber
\end{align}
In terms of the color vector space introduced in \eqref{eq:colorbasis} and of the quantities we have just defined we find the explicit form
\begin{equation}\label{delta_4_main}
\mathbf{\Delta}_4^{(3)} = 8\left({\small
\begin{array}{ccccc}
 -2N_c(2D_1 + D_2 +4 C) &~~ & 2N_c(2D_1 + 3D_2 + 2C)&~~ &  2N^2_c(2D_2- C) \\[8pt]
 2N_c(3D_1 + 2D_2 + 2C) &~~ & -2N_c(D_1 +2 D_2 -4 C) &~~ & 2N^2_c(2D_1- C)   \\[8pt]
D_1 + 2 N_c^+ D_2 - N_c^- C  \; &~~ & 2 N_c^+ D_1 + D_2 -N_c^- C \; &~~ & 6N_c(D_1 + D_2 - C)
\end{array}
}\right),
\end{equation}
where $N_c^{\pm} = (N_c^2\pm1)/2$ and $C = \zeta_5 + 2 \zeta_2 \zeta_3$.
Unlike $ \mathbf{\Gamma}_\text{dipole}$, $\mathbf{\Delta}^{(3)}_4$ does not depend explicitly on the factorization scale $\mu^2$.
We highlight the contributions to the quadrupole soft divergences, and in particular to the colour correlation pattern in the first and second line of eq.~\eqref{eq:quadrupole}, by drawing a couple of  representative diagrams in figure~\ref{diagrams_red}.\\
The coefficients of the perturbative expansion for the finite remainders\begin{equation}
\mathbfcal{H}_{\lambda,\text{fin}}=\sum_{\ell\geq 0}\aso^\ell \mathbfcal{H}^{(\ell)}_{\lambda,\text{fin}}
\end{equation}
can be obtained according to
\begin{align}\label{one}
\mathbfcal{H}_{\lambda,\:\text{fin}}^{(0)} &= \mathbfcal{H}_{\lambda}^{(0)} \;,\nonumber\\
\mathbfcal{H}_{\lambda,\:\text{fin}}^{(1)} &= \mathbfcal{H}_{\lambda,\: \text{ren}}^{(1)} - \mathbfcal{I}_1\mathbfcal{H}_{\lambda,\: \text{ren}}^{(0)} \;,  \nonumber\\
\mathbfcal{H}_{\lambda,\:\text{fin}}^{(2)} &= \mathbfcal{H}_{\lambda,\: \text{ren}}^{(2)} - \mathbfcal{I}_2\mathbfcal{H}_{\lambda,\: \text{ren}}^{(0)} - \mathbfcal{I}_1\mathbfcal{H}_{\lambda,\: \text{ren}}^{(1)} \;,  \nonumber \\
\mathbfcal{H}_{\lambda,\:\text{fin}}^{(3)} &= \mathbfcal{H}_{\lambda,\: \text{ren}}^{(3)} - \mathbfcal{I}_3\mathbfcal{H}_{\lambda,\: \text{ren}}^{(0)} - \mathbfcal{I}_2\mathbfcal{H}_{\lambda,\: \text{ren}}^{(1)} - \mathbfcal{I}_1\mathbfcal{H}_{\lambda,\: \text{ren}}^{(2)} \;,  
\end{align}
with
\begin{equation}\label{eq:I}
\mathbfcal{I}_{1}  = \mathbfcal{Z}_1,\qquad
\mathbfcal{I}_{2} =  \mathbfcal{Z}_2 - \mathbfcal{Z}_1^2,\qquad
\mathbfcal{I}_{3} = \mathbfcal{Z}_3    -   2\mathbfcal{Z}_1  \mathbfcal{Z}_2 + \mathbfcal{Z}_1^3 + \mathbf{\Delta}_4^{(3)}  \,,
\end{equation}
where the $\mathbfcal{Z}_n$ are the coefficients of the expansion of $\mathbfcal{Z}$ in $\aso$ and explicitly read~\cite{Becher:2009qa, Caola:2021rqz}:
\begin{align}
 \mathbfcal{Z}_0 &=  1 \,,  \nonumber \\
 \mathbfcal{Z}_1 &=  \frac{\Gamma'_0}{4 \epsilon^2} + \frac{\mathbf{\Gamma}_0}{2 \epsilon} \,,   \label{eq:Zir}\nonumber\\
\mathbfcal{Z}_2 &=   \frac{{\Gamma_0'}^2}{32 \epsilon^4} + \frac{\Gamma'_0}{8 \epsilon^3} \left( \mathbf{\Gamma}_0 - \frac{3}{2} \beta_0  \right) +  \frac{\mathbf{\Gamma}_0}{8 \epsilon^2}(\mathbf{\Gamma}_0 - 2 \beta_0)  + \frac{\Gamma_1'}{16 \epsilon^2} + \frac{\mathbf{\Gamma}_1}{4 \epsilon}\,, \nonumber\\ 
\mathbfcal{Z}_3 &=  \frac{{\Gamma'_0}^3}{384 \epsilon^6}  + \frac{{\Gamma'_0}^2}{64 \epsilon^5}(\mathbf{\Gamma}_0 \!-\! 3 \beta_0) + \frac{\Gamma_0'}{32 \epsilon^4} \left( \mathbf{\Gamma}_0 \!- \!\frac{4}{3} \beta_0 \right) \left( \mathbf{\Gamma}_0 \!-\! \frac{11}{3} \beta_0 \right)   + \frac{\Gamma_0' \Gamma_1'}{64 \epsilon^4}  \nonumber\\ 
&\quad +\!\frac{\mathbf{\Gamma}_0}{48\epsilon^3}(\mathbf{\Gamma}_0 -\! 2 \beta_0)(\mathbf{\Gamma}_0 - 4 \beta_0)\!+\! \frac{\Gamma'_0}{16 \epsilon^3} \left( \mathbf{\Gamma}_1\!- \!\frac{16}{9} \beta_1\right)  + \frac{\Gamma_1'}{32 \epsilon^3} \left( \mathbf{\Gamma}_0 \!-\! \frac{20}{9} \beta_0 \right)+ \frac{\mathbf{\Gamma}_0 \mathbf{\Gamma}_1}{8 \epsilon^2} \nonumber\\ 
&\quad -\! \frac{\beta_0 \mathbf{\Gamma}_1 + \beta_1 \mathbf{\Gamma}_0}{6 \epsilon^2} + \frac{\Gamma_2'}{36 \epsilon^2 } + \frac{\mathbf{\Gamma}_2 + \mathbf{\Delta}_4^{(3)}}{6 \epsilon} \; .
\end{align}
Above we have used
\begin{equation}
\Gamma'(\aso) = \frac{\partial \mathbf{\Gamma}(\{p\},\aso,\mu)}{\partial \log \mu} =  -\gamma^\text{K} \sum_i C_i  = \sum_{\ell \geq 0} \aso^{\ell+1} \Gamma_\ell' ,
\end{equation}
with the last equal sign giving the definition of the perturbative coefficients $\Gamma_\ell' $.\\
The explicit expression for the perturbative expansions of the cusp anomalous dimension and of the quark (gluon) collinear anomalous dimensions are given in the appendix.

\section{Checks and exact results} \label{Checks}
\noindent

First, we have checked that our results for the lower loop amplitudes are consistent with the literature.
In particular, we have compared our tree-level, one-loop and two-loop results for the bare helicity amplitudes for $q\bar{q}\to gg$ in the helicity configurations~\eqref{eq:refhelicities} against the results provided in the ancillary files of ref.~\cite{Ahmed:2019qtg} and find analytical agreement through to weight six.
We have also checked that our one-loop expressions for $q\bar{q}\to gg$ and $qg\to qg$  match results obtained with the automated one-loop generator \texttt{OpenLoops}~\cite{Cascioli:2011va, Buccioni:2019sur}.
At the three-loop level, we have verified that the IR singularities of our results for the renormalized helicity amplitudes in eq.~\eqref{hel_ampls_ren} match the pattern predicted by eqs.~\eqref{eq:IR_factorization}-\eqref{eq:I},
which provides a highly non-trivial check.
From the high energy limit of our amplitudes we extract the quark and gluon impact factors and find that they are consistent with 
previous results, which tests lower loop contributions to the renormalized amplitude up to weight six.
Moreover, we extract the gluon Regge trajectory and find agreement with previous results, which provides a stringent check of the finite contributions to the three-loop amplitudes presented in this paper.
The high energy limit will be described in more detail in the next section.

Our analytic results for the three-loop finite remainders $ \mathbfcal{H}_{{\lambda},\: \text{fin}} $ are expressed in terms of harmonic polylogarithms with transcendental weight up to six.
Alternatively, these can be converted to a functional basis of logarithms, classical polylogarithms and a few multiple polylogarithms with at most three-fold nested sums~\cite{Caola:2020dfu}.
We provide a general conversion table for harmonic polylogarithms up to weight six in the ancillary files of the \texttt{arXiv} submission of this article. 

From our results for the process $q\bar{q}\to gg$ we also derive explicit expressions for the helicity amplitudes for $qg\to qg$ scattering, which
requires a non-trivial analytical continuation.
Details for this procedure are given in ref.~\cite{Caola:2021rqz}.
The remaining partonic channels 
$gg\to q\bar{q}$ and $g\bar{q} \to g\bar{q}$ are not provided explicitly, since they can be obtained by a simple crossing of external legs without any non-trivial analytic continuation.
While our results are relatively compact, of the order of 1 megabyte per partonic channel, they are too lengthy to be presented here. We include them in computer-readable format in the ancillary files on \texttt{arXiv}.

\begin{figure}[thp]
\centering
\subfigure[]{
  \includegraphics[clip,viewport=0 0 700 500,width=0.485\columnwidth]{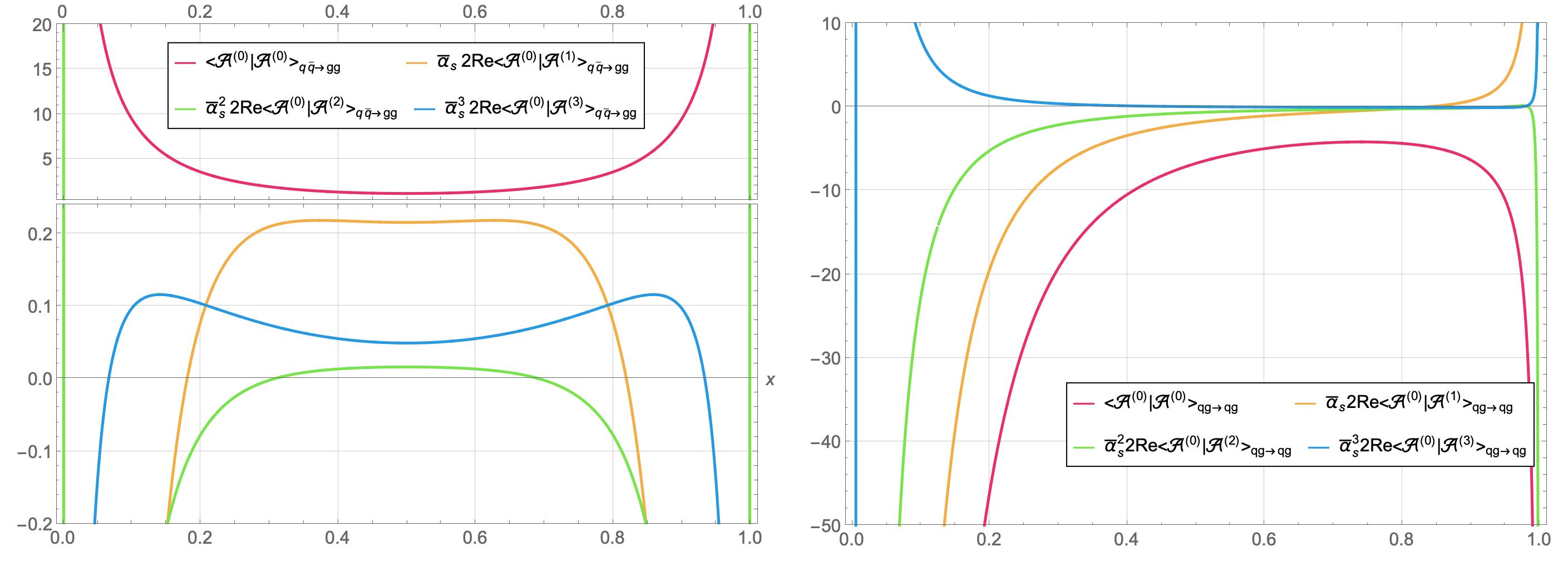}} \label{subfig:a}
\subfigure[]{
  \includegraphics[clip,viewport=700 0 1400 500,width=0.485\columnwidth]{grid_born_projection.jpg}} \label{subfig:b}
\\
\;\subfigure[]{
  \includegraphics[clip,width=0.47\columnwidth]{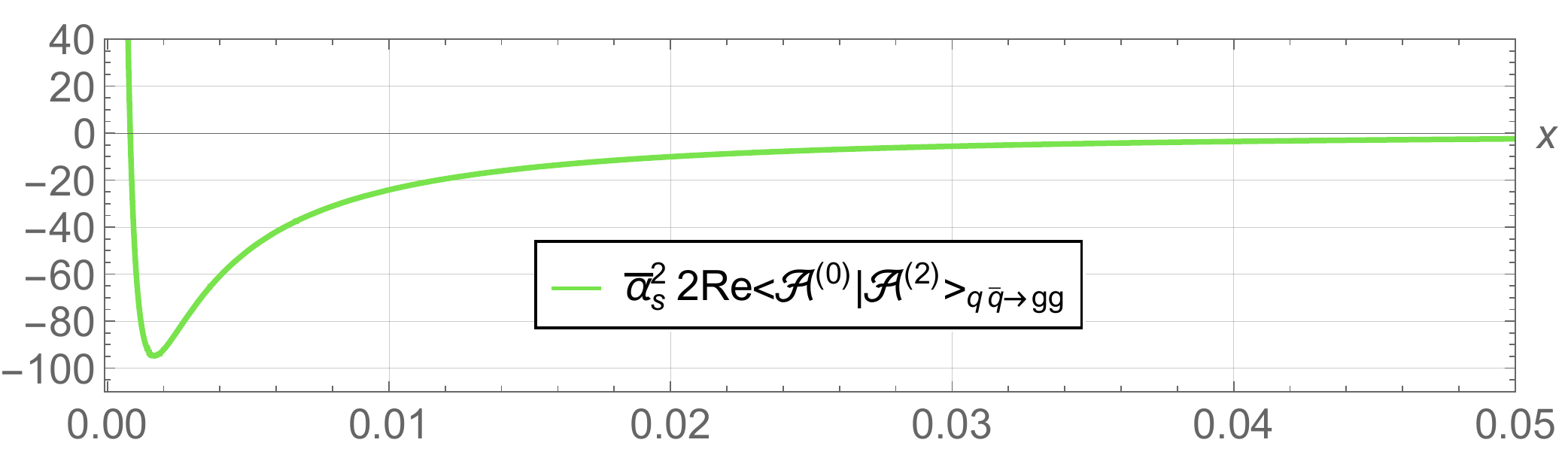}} \label{subfig:c}
\; \subfigure[]{
  \includegraphics[clip,width=0.47\columnwidth]{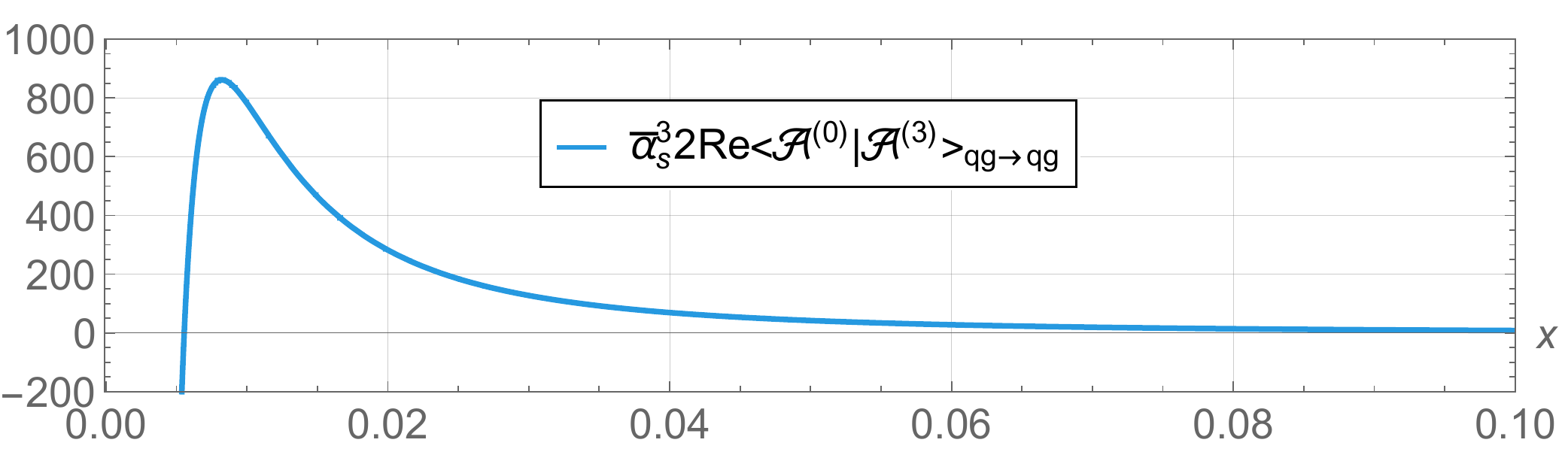}} \label{subfig:d}  
\caption{Perturbative amplitudes up to three loops interfered with the tree-level amplitude for $q\bar{q}\to gg$ (panel~a) and $qg \to qg$ (panel~b) in dependence of $x=-t/s$ . The two-loop contribution to $q\bar{q}\to gg$ diverges to $+\infty$ near $x=0^+,1^-$ (panel~c shows details near $x=0^+$), while the three-loop contribution to $qg \to qg$ diverges to $-\infty$ near $x=0^+$ (panel~d).
}
\label{fig:born}
\end{figure}

\begin{figure}[thp]
    \includegraphics[width=1\linewidth]{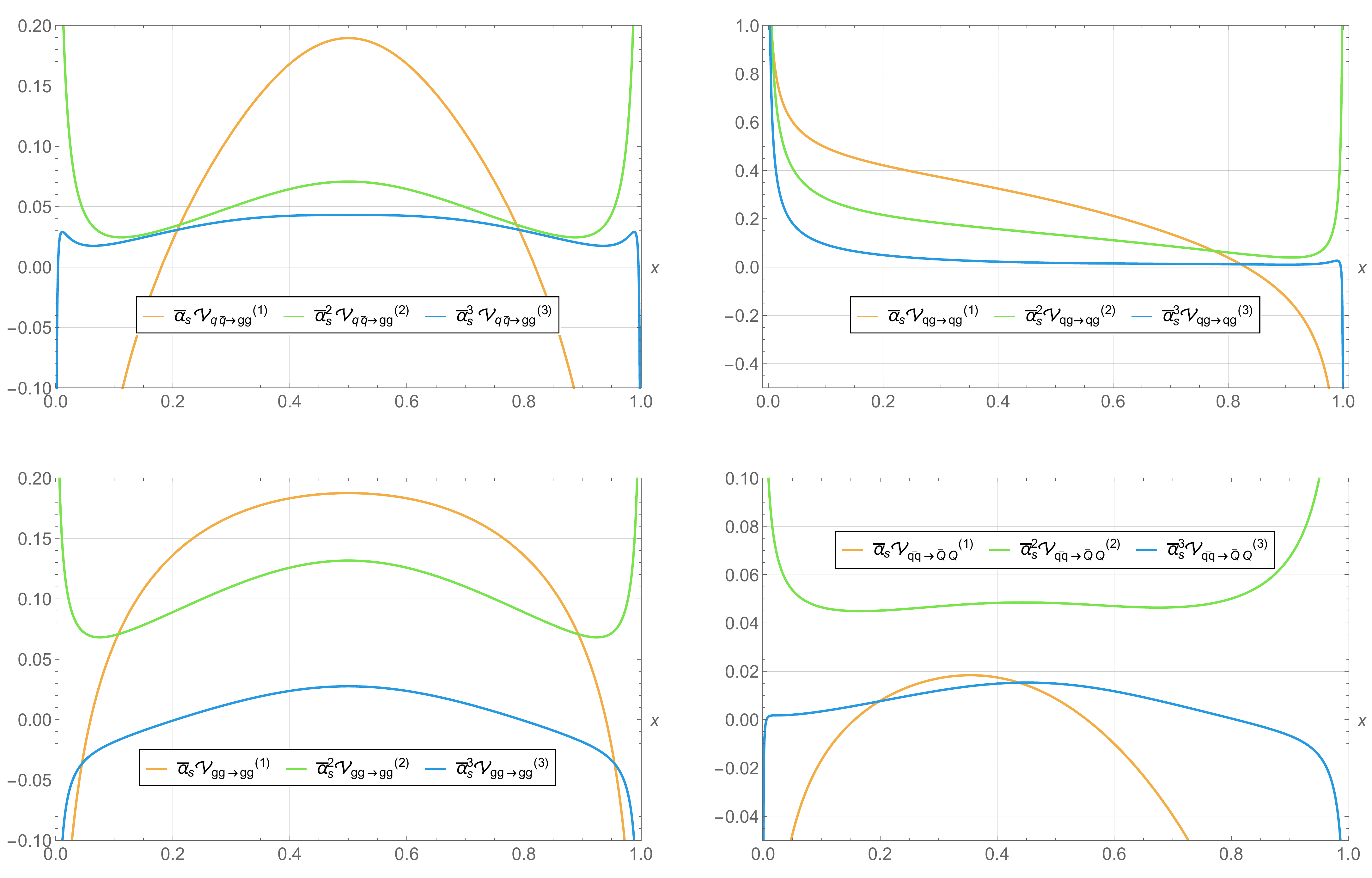}
    \caption{Perturbative expansion of the amplitude squared for the processes $q\bar{q}\to gg$, $qg \to qg$, $gg\to gg$ and  $q\bar{q}\to \bar{Q}Q$ as functions of $x=-t/s$. Values are normalized by the tree-level amplitude squared.
}  \label{fig:full_amplitudes}
\end{figure}

In figure~\ref{fig:born} we show the finite remainder of the amplitude at different loop orders interfered with the tree-level amplitude for the processes $q\bar{q}\to gg$ and $qg \to qg$.
The interferences are averaged (summed) over polarizations and color in the initial (final) state. 
Additionally, since with the results of this paper all $2 \to 2$ partonic channels are now available in three-loop massless QCD, we find it useful to compare virtual corrections for the processes $q\bar{q}\to gg$, $qg \to qg$, $gg\to gg$ and  $q\bar{q}\to \bar{Q}Q$.
In figure~\ref{fig:full_amplitudes}, we show the contributions to the squared amplitude at different orders in $\aso$, normalized by the respective tree-level squared amplitude.
Again, we average (sum) over polarization and color in the initial (final) states.
Below we define more in detail the quantities we present in the plots.

We rewrite the finite amplitude as a vector in color and helicity space
\begin{align} \label{eq:bra-ket_def}
 | \mathcal A\rangle   = 4 \pi \as 
\sum_{\ell\geq 0} \aso^\ell \; | \mathcal A^{(\ell)} \rangle
\end{align}
and define the contraction between different elements in this vector space as
\allowdisplaybreaks[1]
\begin{align}
\langle \mathcal A^{(\ell)} | 
\mathcal A^{(\ell')}\rangle 
\equiv  \mathcal N
\sum_{i,j,\bm \lambda}
\mathcal C_i^\dagger \mathcal C_{j}
|s_{\bm \lambda}|^2
\mathcal H^{[i],(\ell)^*}_{\bm \lambda, \rm fin}
\mathcal H^{[j],(\ell')}_{{\bm \lambda}, \rm fin},
\end{align}
where the factor $4 \pi \as$ in eq.~\eqref{eq:bra-ket_def} replicates the overall normalization of eq.~\eqref{eq:decomp_color}.
$\mathcal N$ is the initial-state color and polarization averaging factor, which depends on the process and takes the following values:
\begin{equation} \label{eq:normalization_factors}
    \mathcal N = 
\begin{cases}
\frac{1}{4N_c^2}  \quad\quad\quad&\text{ for $q\bar{q}\to gg$}, \\
\frac{1}{4N_c(N_c^2-1)}  \quad&\text{ for $qg\to qg$},\\
\frac{1}{4(N_c^2-1)^2}  \quad&\text{ for $gg \to gg$},\\
\frac{1}{4N_c^2}  \quad\quad\quad&\text{ for $q \bar{q}\to \bar{Q}Q $}.
\end{cases}
\end{equation}
The initial and final state polarization sum
runs over all helicity configurations. The color factors $\mathcal{C}_i$ and the spinor factors $s_{\bf \lambda}$ are different for the various processes: for $q\bar{q}\to gg$ they are given in eqs.~\eqref{eq:colorbasis} and \eqref{eq:spinor_factors}, while for $qg\to qg$ they are obtained by applying the transformation $p_2 \leftrightarrow p_3$ to those of $q\bar{q}\to gg$. For the other two channels $gg \to gg$ and $q \bar{q}\to \bar{Q}Q $, they can be found in refs.~\cite{Caola:2021izf} and \cite{Caola:2021rqz} respectively.

We expand the squared amplitude normalized by the tree-level contribution in $\aso$ according to
\begin{align}
\frac{\langle \mathcal A | 
\mathcal A\rangle}{\langle \mathcal A^{(0)} | 
\mathcal A^{(0)}\rangle}  =  
\mathcal{V}^{(0)} 
+
\aso\mathcal{V}^{(1)} 
+
\aso^2\mathcal{V}^{(2)} 
+
\aso^3\mathcal{V}^{(3)} 
+
O(\aso^4)
\end{align}
with
\begin{align}
\mathcal{V}^{(0)} &=
1, \quad
&\mathcal{V}^{(1)} &= 2\frac{\myRe\langle \mathcal A^{(0)}
| \mathcal A^{(1)}\rangle}{\langle \mathcal A^{(0)} | 
\mathcal A^{(0)}\rangle}, \quad \notag\\
\mathcal{V}^{(2)} &= \frac{\langle \mathcal A^{(1)} | 
\mathcal A^{(1)}\rangle}{\langle \mathcal A^{(0)} | 
\mathcal A^{(0)}\rangle} 
+
2\frac{\myRe\langle \mathcal A^{(0)} | 
\mathcal A^{(2)}\rangle}{\langle \mathcal A^{(0)} | 
\mathcal A^{(0)}\rangle}, \quad
&\mathcal{V}^{(3)} &=  2\frac{\myRe \langle \mathcal A^{(1)} | 
\mathcal A^{(2)}\rangle}{\langle \mathcal A^{(0)} | 
\mathcal A^{(0)}\rangle} 
+
2 \frac{\myRe\langle \mathcal A^{(0)} | 
\mathcal A^{(3)}\rangle}{\langle \mathcal A^{(0)} | 
\mathcal A^{(0)}\rangle} .
\end{align}
Finally, for the numerical evaluation, we have set $\mu^2=s = m_Z^2$, $\alpha_s(\mu)=0.118$, $n_f=5$ and $N_c=3$.

\section{High energy limit} \label{HighEn}
\noindent
In the high-energy or \emph{Regge limit}, quantum field theoretic scattering amplitudes become particularly simple and are known to exhibit universal factorization properties.
In the following, we consider the process
\begin{equation}\label{eq:channelqg}
    q(p_1)\;+ \;g(p_2) \; \to \; q(p_3)\;+ \;g(p_4),
\end{equation}
for which $t$-channel gluon exchanges provide the dominant contribution to the amplitude at high energies.
The Regge limit is defined as $s\to \infty$ for fixed scattering angle, that is, $|s| \approx |u| \gg |t|$,
where $s=(p_1+p_2)^2$, $t=(p_1-p_3)^2$, $u=-s-t$ in terms of the momenta in~\eqref{eq:channelqg}.
For the variable $x=-t/s$, the Regge limit corresponds to $x\to 0$.

Following the investigation~\cite{Caron-Huot:2017fxr,Collins:1977jy}, we split the renormalized amplitude into the definite $s \leftrightarrow u$ signature component
\begin{equation}
  \mathbfcal{H}_{\mathrm{qg \rightarrow qg,\pm}} = \frac{1}{2}\left.[ \mathbfcal{H}_{\mathrm{qg \rightarrow qg}}(s,u)   \pm   \mathbfcal{H}_{\mathrm{qg \rightarrow qg}}(u,s)  \right]  \,.
\end{equation}
The definite-signature amplitudes $\mathbfcal{H}_{\mathrm{qg \rightarrow qg},+}$ and $\mathbfcal{H}_{\mathrm{qg \rightarrow qg},-}$ are referred to as the even and odd amplitudes.
We expand them up to third order in $\aso$,
\begin{align}\label{eq:Regge_expand}
 \mathbfcal{H}_{\mathrm{qg \rightarrow qg},\pm}   &=  \sum_{\ell=0}^3 \aso^\ell \sum_{k=0}^\ell L^k \mathbfcal{H}_{\mathrm{qg \rightarrow qg}} ^{(\pm,\ell,k)}, 
\end{align}
where we use for the signature-symmetric logarithm
\begin{equation}
L =  -\ln(x) -  \frac{i\pi}{2} \approx \frac{1}{2}  \left[ \ln\left( \frac{-s-i \delta }{-t}  \right) + \ln\left( \frac{-u-i \delta }{-t}  \right)  \right]
\end{equation}
and the color operators~\cite{DelDuca:2013ara,DelDuca:2014cya} are
\begin{align}
&  \mathbf{T}_s^2 = (\mathbf{T}_1 \!+\! \mathbf{T}_2)^a(\mathbf{T}_1\!+\! \mathbf{T}_2)^a ,  \;\quad \mathbf{T}_t^2 = (\mathbf{T}_1 \!+\! \mathbf{T}_3)^a(\mathbf{T}_1 \!+\! \mathbf{T}_3)^a,  \nonumber \\
& \mathbf{T}_u^2 = (\mathbf{T}_1 \!+\! \mathbf{T}_4)^a(\mathbf{T}_1 \!+\! \mathbf{T}_4)^a ,\;\quad \mathbf{T}_{s-u}^2 = \frac{1}{2}(\mathbf{T}_s^2 -\mathbf{T}_u^2 ).
\end{align}
Here the $\mathbf{T}_i$ (i=1,\ldots,4) are assigned according to eq.~\eqref{convention}. Explicitly, we find
\begin{align}
\Scale[0.95]{
   \mathbf{T}_s^2 \!=\!   \left(
\begin{array}{ccc}
C_A\!+\!C_F & 0 & 2 \\
 0 & C_F & \shortminus2 \\
 1/2 & 0 &  C_A\!+\!C_F \\
\end{array}
\right), 
\;
\mathbf{T}_t^2 \!=\!  \left(
\begin{array}{ccc}
 C_A & 0 & 0 \\
 0 & C_A & 0 \\
 \shortminus1/2 & \shortminus1/2 & 0 \\
\end{array}
\right),
\; 
\mathbf{T}_u^2\! =\!  
\left(
\begin{array}{ccc}
 C_F & 0 & \shortminus 2 \\
 0 & C_A\!+\!C_F & 2 \\
 0 & 1/2 & C_A\!+\!C_F\
\end{array}
\right).
}
\end{align}
Following ref.~\cite{Caron-Huot:2017fxr}, one can show that the coefficients $\mathbfcal{H}_{\mathrm{qg \rightarrow qg}} ^{(-,\ell,k)}$($\mathbfcal{H}_{\mathrm{qg \rightarrow qg}} ^{(+,\ell,k)}$) are purely imaginary(real). The $t$-channel exchange of an even number of Reggeons contributes only to $\mathbfcal{H}_{\mathrm{qg \rightarrow qg}} ^{(+,\ell,k)}$, while the $t$-channel exchange of an odd number of Reggeons contributes only to $\mathbfcal{H}_{\mathrm{\mathrm{qg \rightarrow qg}}}  ^{(-,\ell,k)}$.
A single Reggeon exchange contributes to the \emph{Regge pole} contribution, while a multiple Reggeon exchange in general can have non-vanishing contributions to both Regge pole and \emph{Regge cuts}~\cite{PhysRev.137.B949,Collins:1977jy,Falcioni:2021dgr,Falcioni:2021buo}.
Up to next-to-leading logarithmic (NLL) accuracy, the odd signature amplitude is completely determined by the gluon Regge trajectory and by the so-called quark and gluon impact factors, that describe the interaction of the reggeized gluon with external states. The factorization structure for the odd amplitude becomes more complex in the next-to-next-to-leading logarithmic (NNLL) approximation, as both  Regge pole and  Regge cut~\cite{Fadin:2020lam,Caron-Huot:2017fxr,DelDuca:2001gu,DelDuca:2013ara} contribute at this order. For the even amplitude, only the Regge cut contributes at the NLL level~\cite{Caron-Huot:2017fxr} and breaks the simple exponential structure already at this logarithmic order.
Starting from NNLL, the odd-signature amplitude receives contributions from both Regge pole and Regge cuts. In ref.~\cite{Falcioni:2021buo}, a scheme has been proposed to disentangle the two.
As in our previous paper \cite{Caola:2021izf}, we adopt this scheme to study the high-energy behaviour of $qg \to qg$ to three loops up to NNLL. 

Following the framework outlined in~\cite{Falcioni:2021buo}, we assume that, by setting the renormalization scale to $\mu^2 = -t$, eq.~\eqref{eq:Regge_expand} can be written as
\begin{align}\label{eq:Regge_L_expansion}
 \mathbfcal{H}_{\mathrm{qg \rightarrow qg},\pm}   &= \: Z_q\;Z_g  \:e^{L\mathbf{T}_t^2 \tau_g} \sum_{\ell=0}^3 \aso^\ell \sum_{k=0}^\ell L^k \mathbfcal{O}^{\pm,(\ell)}_k  \mathbfcal{H}_{\mathrm{qg \rightarrow qg}} ^{(0)}, 
\end{align}
where $\tau_g = \sum_{\ell=1} \aso^\ell \tau_\ell $ is the gluon Regge trajectory and the factors  $Z_q = \sum_{\ell=0} \aso^\ell Z_q^{(\ell)}$ and $Z_g = \sum_{\ell=0} \aso^\ell Z_g^{(\ell)}$ capture the collinear poles of the amplitude~\cite{Caron-Huot:2017fxr} for quarks and gluons, respectively.
Up to $O(\aso)$ we have
\begin{align}
 Z_i^{(0)} & = 1 \, , \nonumber\\
 Z_i^{(1)} & = -C_i \gamma^\text{K}_1 \frac{1}{ \ep^2} + 4\gamma_1^i\frac{1}{\ep} \, , \nonumber \\
 Z_i^{(2)} & = C_i^2  \frac{(\gamma^\text{K}_1)^2}{2 \ep^4} + C_i\left[\frac{1}{\ep^3}\gamma^\text{K}_1\left(\frac{3 \beta_0}{4} - 4\gamma_1^i \right) - \frac{\gamma^\text{K}_2}{\ep^2}  \right] + \frac{2}{\ep^2}\gamma_1^i\left(4\gamma_1^i -\beta_0 \right) + \frac{8\gamma_2^i}{ \ep}\, .
\end{align}
The odd signature color operators $\mathbfcal{O}^{-,(\ell)}_k$ contributing at NNLL~\cite{Caron-Huot:2017fxr} are
\begin{align}
\label{eq:Regge_odd_operators}
  \mathbfcal{O}^{-,(0)}_0 &=1, \\
\mathbfcal{O}^{-,(1)}_0 
&= \mathcal{I}^q_1+\mathcal{I}^g_1,  \nonumber\\
 \mathbfcal{O}^{-,(2)}_0 
&= \left[ \mathcal{I}^q_2 +\mathcal{I}^g_2 + \mathcal{I}^q_1\mathcal{I}^g_1\right] + \mathcal{B}^{-,\Scale[0.7]{(2)}} [ (\mathbf{T}^2_{s-u} )^2 - N_c^2/4 
], \nonumber\\
 \mathbfcal{O}^{-,(3)}_1 &=  \mathcal{B}_1^{-,\Scale[0.7]{(3)}}\mathbf{T}^2_{s-u}[\mathbf{T}^2_{t},\mathbf{T}^2_{s-u}] 
+ \mathcal{B}_2^{-,\Scale[0.7]{(3)}}[\mathbf{T}^2_{t},\mathbf{T}^2_{s-u}] \mathbf{T}^2_{s-u} ,  \nonumber
\intertext{
and the even signature ones contributing at NLL~\cite{Caron-Huot:2017fxr} are
}
   \mathbfcal{O}^{+,(1)}_0 &= i \pi \,\mathcal{B}^{+,(1)} \, \mathbf{T}_{s-u}^2, \; 
   \nonumber\\
   \mathbfcal{O}^{+,(2)}_1 &= i \pi \, \mathcal{B}^{+,(2)}  \,[ \mathbf{T}_{t}^2, \mathbf{T}_{s-u}^2],  \nonumber \\
 \mathbfcal{O}^{+,(3)}_2 &= i \pi \, \mathcal{B}^{+,(3)} \, [\mathbf{T}_{t}^2,[ \mathbf{T}_{t}^2, \mathbf{T}_{s-u}^2]] . \label{eq:Regge_even_operators} 
\intertext{
The coefficients $\mathcal{B}^{\pm,(\ell)}$ describe 
the process independent Regge cut contributions~\cite{Caron-Huot:2013fea,Caron-Huot:2017fxr,Falcioni:2021buo} and we report them below for convenience.
The odd-signature ones are
}
\mathcal{B}^{-,(2)}& = \frac{2\pi^2}{3} r_\Gamma^2 \left( \frac{3}{\ep^2} - 18 \ep \zeta_3 - 27 \ep^2 \zeta_4+ \mathcal{O}(\ep) \right),\nonumber \\
\mathcal{B}_1^{-,(3)} & = 64\pi^2 r_\Gamma^3  \left( \frac{1}{48\ep^2} + \frac{37}{24} \zeta_3 + \mathcal{O}(\ep) \right) ,\nonumber \\
\mathcal{B}_2^{-,(3)} &= 64\pi^2 r_\Gamma^3  \left( \frac{1}{24\ep^2} + \frac{1}{12} \zeta_3 + \mathcal{O}(\ep) \right),
\intertext{
while for even signature one finds
}
   \mathcal{B}^{+,(1)} &= \, r_{\Gamma} \: \frac{2}{\epsilon},  \nonumber\\
 \mathcal{B}^{+,(2)} &= - \frac{r_{\Gamma}^2}{2} \left(  \frac{4}{\epsilon^2} +
 72 \zeta_3 \epsilon + 108 \zeta_4 \epsilon^2 +\mathcal{O}(\epsilon^3)\right),  \nonumber \\
 \mathcal{B}^{+,(3)} &= \frac{r_{\Gamma}^3}{6} \bigg(  \frac{8}{\epsilon^3}  - 176 \zeta_3 - 264 \zeta_4 \epsilon  -  5712 \zeta_5 \epsilon^2  + \mathcal{O}(\epsilon^3) \bigg) . 
\end{align}
$\mathcal I^q_\ell$ and $\mathcal I^g_\ell$ are
the perturbative expansion coefficients of the quark and gluon impact factors;
they can be extracted from the one- and two-loop calculation~\cite{Ahmed:2019qtg}. The explicit expressions are rather long and are reported to the required orders in $\epsilon$ in appendix \ref{app:impact}.\\
With the perturbative expansion of $\tau_g$ up to the three-loop order obtained in \cite{Caola:2021izf} (and provided in appendix \ref{app:impact}), we have all the ingredients to fully predict the Regge limit of the process $qg\to qg$ through eq.~\eqref{eq:Regge_L_expansion}, which only requires the tree-level amplitude $\mathcal{H}_{\mathrm{qg \rightarrow qg}} ^{(0)}$ as an input.

We find by explicit calculation that the high energy limit of our results for the $qg\to qg$ three-loop amplitude indeed agrees with this prediction and confirms in particular the literature results \cite{Caron-Huot:2017fxr,Falcioni:2021dgr,Fadin:1996tb,Blumlein:1998ib,DelDuca:2021vjq} for the gluon Regge trajectory as well as quark and gluon impact factors in QCD.
This provides a highly non-trivial test of the universality of high energy factorization in QCD.

\section{Conclusions} \label{Conc}
\noindent
In this paper, we have presented the three-loop helicity amplitudes for
quark-gluon scattering processes in full-color, massless QCD. To perform this calculation, we have made use of various cutting-edge techniques, in particular to handle the Lorentz decomposition of the scattering amplitude and to solve the highly non-trivial system of integration-by-parts identities required to reduce the amplitude to master integrals.

In addition to our previous calculations for the scattering of four quarks and of four gluons, these
latest analytical results confirm predictions for the infrared poles of four-point amplitudes in QCD, also for processes with external states in different color representations.
Moreover, our results have made it possible to verify the factorization properties of partonic amplitudes in the Regge limit.
With this work, all three-loop amplitudes for parton-parton scattering processes are publicly available, providing the virtual corrections to dijet production at N$^3$LO.
\newline

\acknowledgements
 The research of FC was supported by the ERC Starting Grant 804394 \textsc{hipQCD} and by the UK Science and Technology Facilities Council (STFC) under grant ST/T000864/1. GG was supported by the Royal Society grant URF/R1/191125. AvM was supported in part by the National Science Foundation through Grant 2013859. LT was supported by the Excellence Cluster ORIGINS funded by the Deutsche Forschungsgemeinschaft (DFG, German Research Foundation) under Germany’s Excellence Strategy - EXC-2094 - 390783311, by the ERC Starting Grant 949279 HighPHun and, in the initial phase of this work, by the Royal Society through grant URF/R1/191125.

\bibliography{biblio}

\newpage
\onecolumngrid
\appendix
\appendix

\makeatletter
\renewcommand\@biblabel[1]{[#1S]}
\makeatother

\allowdisplaybreaks[1]

\section{Anomalous dimensions}\label{app:andim}
In this appendix, we list the perturbative expansions of the cusp anomalous dimension and  of the quark and gluon collinear anomalous dimensions,
\begin{equation}
\gamma^{\text{K}} = \sum\limits_{n=0} \left(\frac{\as}{4 \pi}\right)^{n+1}  \gamma_n^\text{K} ,  \qquad \gamma^{g/q} = \sum\limits_{n=0} \left(\frac{\as}{4 \pi}\right)^{n+1}  \gamma_n^{q/g}.
\end{equation}
The required expansion coefficients of the cusp anomalous dimension read~\cite{Korchemsky:1987wg,Moch:2004pa,Vogt:2004mw}
\begin{align}\label{K}
   \gamma_0^\text{K} &= 4 \,, \nonumber\\
   \gamma_1^\text{K} &= \left( \frac{268}{9} 
    - \frac{4\pi^2}{3} \right) C_A - \frac{40}{9}\, n_f \,,
  \nonumber\\
   \gamma_2^\text{K} &= C_A^2 \left( \frac{490}{3} 
    - \frac{536\pi^2}{27}
    + \frac{44\pi^4}{45} + \frac{88}{3}\,\zeta_3 \right) + C_A  n_f  \left(  \frac{80\pi^2}{27}- \frac{836}{27} - \frac{112}{3}\,\zeta_3 \right) \nonumber\\
    & \quad + C_F n_f \left(32\zeta_3 - \frac{110}{3}\right) - \frac{16}{27}\, n_f^2 \; .  
\intertext{
The required expansion coefficients of the quark collinear anomalous dimension are~\cite{Moch:2005id} }
   \gamma_0^q &= -3 C_F \,, \nonumber\\
   \gamma_1^q &= C_F^2 \left( -\frac{3}{2} + 2\pi^2
    - 24\zeta_3 \right)
    + C_F C_A \left( - \frac{961}{54} - \frac{11\pi^2}{6} 
    + 26\zeta_3 \right)
    + C_F  n_f \left( \frac{65}{27} + \frac{\pi^2}{3} \right) ,
    \nonumber\\
   \gamma_2^q &= C_F^3 \left( -\frac{29}{2} - 3\pi^2
    - \frac{8\pi^4}{5}
    - 68\zeta_3 + \frac{16\pi^2}{3}\,\zeta_3 + 240\zeta_5 \right) 
    \nonumber\\
   &\mbox{}+ C_F^2 C_A \left( - \frac{151}{4} + \frac{205\pi^2}{9}
    + \frac{247\pi^4}{135} - \frac{844}{3}\,\zeta_3
    - \frac{8\pi^2}{3}\,\zeta_3 - 120\zeta_5 \right) \nonumber\\
   &\mbox{}+ C_F C_A^2 \left( - \frac{139345}{2916} - \frac{7163\pi^2}{486}
    - \frac{83\pi^4}{90} + \frac{3526}{9}\,\zeta_3
    - \frac{44\pi^2}{9}\,\zeta_3 - 136\zeta_5 \right) \nonumber\\
   &\mbox{}+ C_F^2  n_f \left( \frac{2953}{54} - \frac{13\pi^2}{9} 
    - \frac{14\pi^4}{27} + \frac{256}{9}\,\zeta_3 \right) 
    \nonumber\\
   &\mbox{}+ C_F C_A n_f \left( - \frac{8659}{729}
    + \frac{1297\pi^2}{243} + \frac{11\pi^4}{45} 
    - \frac{964}{27}\,\zeta_3 \right) \nonumber\\
   &\mbox{}+ C_F  n_f^2 \left( \frac{2417}{729} 
    - \frac{10\pi^2}{27} - \frac{8}{27}\,\zeta_3 \right) \; ,\\
\intertext{while for the gluon collinear anomalous dimension~\cite{Moch:2005tm} they read
}
   \gamma_0^g &=   -\beta_0   \,, \nonumber\\
   \gamma_1^g  &=    C_A^2 \left( -\frac{692}{27} + \frac{11}{3} \zeta_2 + 2 \zeta_3 \right)+C_A n_f \left( \frac{128}{27} - \frac{2}{3} \zeta_2 \right) + 2 C_F n_f \,,  \nonumber\\
   \gamma_2^g &=   C_A^3\left(\frac{-97186}{729} + \frac{6109}{81} \zeta_2 + \frac{122}{3} \zeta_3- \frac{319}{3} \zeta_4 - \frac{40}{3} \zeta_2 \zeta_3\;- 16 \zeta_5 \right)\nonumber\\
   &\mbox{}+C_A^2 n_f \left( \frac{30715}{1458} - \frac{1198}{81}\zeta_2 + \frac{356}{27} \zeta_3 + \frac{82}{3} \zeta_4 \right) - \frac{11}{9} C_F n_f^2 - C_F^2 n_f \nonumber\\
&\mbox{}+C_A C_F n_f \left( \frac{1217}{27} - 2 \zeta_2 - \frac{152}{9}\zeta_3 - 8 \zeta_4 \right) + C_A n_f^2 \left(-\frac{269}{1458} + \frac{20}{27} \zeta_2 - \frac{56}{27} \zeta_3 \right).
\end{align}

\section{Impact factors and gluon Regge trajectory}
\label{app:impact}
In this appendix we provide expressions relevant for the high-energy limit of the three-loop amplitude discussed in the main text.
The expansion coefficients for the quark and gluon impact factors up to two loops read
\begin{align}
\mathcal{I}^q_1 & =\frac{4-\frac{
   \zeta_2}{2}}{N_c}+N_c \left(\frac{7
   \zeta_2}{2}+\frac{13}{18}\right)-\frac{5 n_f}{9}\nonumber\\
&\quad+\epsilon  \bigg[N_c
   \left(-\frac{\zeta_2}{6}+\frac{10 \zeta_3}{3}+\frac{40}{27}\right)+\frac{1}{N_c}\left(-\frac{3
   \zeta_2}{4}-\frac{7 \zeta_3}{3}+8\right)+n_f
   \left(\frac{\zeta_2}{6}-\frac{28}{27}\right)\bigg]\nonumber\\
&\quad +\epsilon ^2 \bigg[N_c \left(-\frac{13
   \zeta_2}{36}+\frac{35 \zeta_4}{16}-\frac{7 \zeta_3}{9}+\frac{242}{81}\right)+\frac{1}{N_c}\left(-2
   \zeta_2-\frac{47 \zeta_4}{16}-\frac{7 \zeta_3}{2}+16\right)\nonumber\\
&\quad\quad\quad+n_f \left(\frac{5
   \zeta_2}{18}+\frac{7 \zeta_3}{9}-\frac{164}{81}\right)\bigg]\nonumber\\
&\quad +\epsilon ^3 \bigg[N_c \left(-\frac{26 \zeta_2 \zeta_3}{3}-\frac{20 \zeta_2}{27}-\frac{47 \zeta_4}{48}+\frac{36
   \zeta_5}{5}-\frac{91 \zeta_3}{54}+\frac{1456}{243}\right)\nonumber\\
&\quad\quad\quad+\frac{1}{N_c}\left(\frac{7 \zeta_2 \zeta_3}{6}-4 \zeta_2-\frac{141
   \zeta_4}{32}-\frac{31 \zeta_5}{5}-\frac{28 \zeta_3}{3}+32\right)\nonumber\\
&\quad\quad\quad+n_f \left(\frac{14
   \zeta_2}{27}+\frac{47 \zeta_4}{48}+\frac{35 \zeta_3}{27}-\frac{976}{243}\right)\bigg]\nonumber\\
&\quad +\epsilon ^4 \bigg[\frac{1}{N_c}\left(\frac{7 \zeta_2 \zeta_3}{4}-8
   \zeta_2-\frac{47 \zeta_4}{4}-\frac{93 \zeta_5}{10}+\frac{49
   \zeta_3^2}{18}-\frac{56 \zeta_3}{3}-\frac{949 \pi
   ^6}{120960}+64\right) \nonumber\\
&\quad\quad\quad+N_c \left(\frac{7 \zeta_2 \zeta_3}{18}-\frac{121 \zeta_2}{81}-\frac{611 \zeta_4}{288}-\frac{31
   \zeta_5}{15}-\frac{91 \zeta_3^2}{18}-\frac{280 \zeta_3}{81}-\frac{977 \pi ^6}{120960}+\frac{8744}{729}\right)\nonumber\\
&\quad\quad\quad+n_f
   \left(-\frac{7 \zeta_2 \zeta_3}{18}+\frac{82
   \zeta_2}{81}+\frac{235 \zeta_4}{144}+\frac{31 \zeta_5}{15}+\frac{196 \zeta_3}{81}-\frac{5840}{729}\right)\bigg]  + \mathcal{O}(\ep^5)\, ,\label{eq:impact_quark_1}
\\
\mathcal{I}^q_2 & =-\frac{3 N_c^2
   \zeta_2}{2 \epsilon ^2} + N_c^2 \left(\frac{87 \zeta_2}{4}+\frac{25
   \zeta_4}{16}+\frac{41 \zeta
   _3}{9}+\frac{22537}{2592}\right)+\frac{1}{N_c^2}\left(\frac{21
   \zeta_2}{4}-\frac{83 \zeta_4}{16}-\frac{15 \zeta
   _3}{2}+\frac{255}{32}\right)\nonumber\\
   &\quad\quad\quad+N_c n_f \left(-4
   \zeta_2-\frac{23 \zeta
   _3}{9}-\frac{650}{81}\right)+\frac{n_f}{N_c}
   \left(-\zeta_2-\frac{19 \zeta
   _3}{9}-\frac{505}{81}\right)+\frac{25
   n_f^2}{54}+\frac{19 \zeta_2}{2}\nonumber\\
   &\quad\quad\quad-\frac{47
   \zeta_4}{8}-\frac{205 \zeta_3}{18}+\frac{28787}{648} \nonumber\\
&\quad+\epsilon \bigg[ N_c^2 \left(\frac{161 \zeta_2 \zeta_3}{6}+\frac{4055
   \zeta_2}{144}+\frac{587 \zeta_4}{12}+\frac{49 \zeta
   _5}{2}+\frac{898 \zeta
   _3}{27}+\frac{911797}{15552}\right)+n_f^2
   \left(\frac{140}{81}-\frac{5 \zeta_2}{18}\right)\nonumber\\
   &\quad\quad\quad+\frac{1}{N_c^2}\left(\frac{49 \zeta_2 \zeta
   _3}{6}+\frac{325 \zeta_2}{16}-\frac{201 \zeta_4}{16}-3 \zeta
   _5-\frac{166 \zeta_3}{3}+\frac{2157}{64}\right)\nonumber\\
   &\quad\quad\quad+N_c
   n_f \left(-\frac{61 \zeta_2}{36}-\frac{247
   \zeta_4}{24}-\frac{85 \zeta
   _3}{27}-\frac{36031}{972}\right)-\frac{5507 \zeta
   _3}{54}+\frac{746543}{3888}\nonumber\\
   &\quad\quad\quad+\frac{n_f}{N_c} \left(-\frac{13
   \zeta_2}{4}-\frac{83 \zeta_4}{24}-\frac{17 \zeta
   _3}{27}-\frac{11983}{486}\right)+13 \zeta_2
   \zeta_3+\frac{115 \zeta_2}{8}-\frac{1283
   \zeta_4}{48}+\frac{121 \zeta_5}{2} \bigg]\nonumber\\
   &\quad+\epsilon^2 \bigg[N_c^2 \left(-\frac{3613 \zeta_2 \zeta_3}{18}+\frac{5131
   \zeta_2}{864}+\frac{31811 \zeta_4}{288}+\frac{94 \zeta
   _5}{5}-\frac{293 \zeta_3^2}{18}+\frac{12007 \zeta
   _3}{648}+\frac{3251 \pi
   ^6}{120960}\right.\nonumber\\
   &\quad\quad\quad\left.+\frac{23246941}{93312}\right)+N_c n_f
   \left(\frac{625 \zeta_2 \zeta_3}{18}+\frac{1475
   \zeta_2}{108}-\frac{779 \zeta_4}{72}-\frac{143 \zeta
   _5}{5}+\frac{1993 \zeta
   _3}{81}-\frac{805855}{5832}\right)\nonumber\\
   &\quad\quad\quad+\frac{1}{N_c^2}\left(10 \zeta_2 \zeta
   _3+\frac{2287 \zeta_2}{32}-\frac{5627 \zeta_4}{64}-\frac{9
   \zeta_5}{2}+\frac{1255 \zeta_3^2}{18}-\frac{6205 \zeta
   _3}{24}+\frac{7193 \pi
   ^6}{120960}+\frac{13575}{128}\right)\nonumber\\
   &\quad\quad\quad+\frac{n_f}{N_c}\left(\frac{31
   \zeta_2 \zeta_3}{9}-\frac{45 \zeta_2}{4}-\frac{503
   \zeta_4}{144}-\frac{151 \zeta_5}{15}+\frac{623 \zeta
   _3}{81}-\frac{227023}{2916}\right)\nonumber\\
   &\quad\quad\quad+n_f^2
   \left(-\frac{53 \zeta_2}{54}+\frac{5 \zeta_4}{48}-\frac{35 \zeta
   _3}{27}+\frac{404}{81}\right)+\frac{1613 \zeta_2 \zeta
   _3}{36}+\frac{197 \zeta_2}{24}-\frac{27175
   \zeta_4}{144}+\frac{791 \zeta_5}{30}\nonumber\\
   &\quad\quad\quad+\frac{1621 \zeta
   _3^2}{18}-\frac{170951 \zeta_3}{324}+\frac{17 \pi
   ^6}{70}+\frac{16114247}{23328} \bigg]+ \mathcal{O}(\ep^3) \, ,\label{eq:impact_quark_2}
\\
\intertext{and}
\mathcal{I}^g_1 & =   N_c
   \left(4 \zeta_2-\frac{67}{18}\right)+\frac{5 n_f}{9}+\epsilon 
   \Bigg[N_c \left(\frac{17 \zeta_3}{3}+\frac{11 \zeta_2}{12}-\frac{202}{27}\right)+n_f \left(-\frac{\zeta_2}{6}+\frac{28}{27}\right)\Bigg]\nonumber\\
    &\quad+\epsilon ^2 \Bigg[N_c \left(\frac{41
   \zeta_4}{8}+\frac{77 \zeta_3}{18}+\frac{67 \zeta_2}{36}-\frac{1214}{81}\right)+n_f \left(-\frac{7 \zeta_3}{9}-\frac{5 \zeta_2}{18}+\frac{164}{81}\right)\Bigg]\nonumber\\
    &\quad+\epsilon ^3 \Bigg[N_c \left(-\frac{59 \zeta_2 \zeta_3}{6}+\frac{67 \zeta_5}{5}+\frac{517 \zeta_4}{96}+\frac{469 \zeta_3}{54}+\frac{101 \zeta_2}{27}-\frac{7288}{243}\right)\nonumber\\
   &\quad\quad\quad+n_f \left(-\frac{47 \zeta_4}{48}-\frac{35 \zeta_3}{27}-\frac{14
   \zeta_2}{27}+\frac{976}{243}\right)\Bigg]\nonumber\\
   &\quad+\epsilon ^4 \Bigg[N_c \left(-\frac{\pi ^6}{4320}-\frac{70 \zeta_3^2}{9}-\frac{77 \zeta_2 \zeta_3}{36}+\frac{341 \zeta_5}{30}+\frac{3149 \zeta_4}{288}+\frac{1414 \zeta_3}{81}+\frac{607 \zeta_2}{81}-\frac{43736}{729}\right)\nonumber\\
&\quad\quad\quad+n_f \left(\frac{7 \zeta_2 \zeta_3}{18}-\frac{31 \zeta_5}{15}-\frac{235
   \zeta_4}{144}-\frac{196 \zeta_3}{81}-\frac{82 \zeta_2}{81}+\frac{5840}{729}\right)\Bigg] + \mathcal{O}(\ep^5)\,,  \label{eq:impact_gluon_1}
\\
\mathcal{I}^g_2 & = -\frac{3 N_c^2 \zeta_2}{2 \epsilon ^2}+N_c^2 \left(\frac{9 \zeta_4}{4}+\frac{88 \zeta_3}{9}+\frac{335
   \zeta_2}{18}-\frac{26675}{648}\right)+N_c n_f
   \left(\frac{2 \zeta_3}{9}-\frac{25 \zeta_2}{9}+\frac{2063}{216}\right)\nonumber\\
   & \quad\quad\quad+\frac{n_f}{N_c} \left(2 \zeta_3-\frac{55}{24}\right)-\frac{25
   n_f^2}{162}\nonumber\\
    &\quad+\epsilon  \Bigg[N_c^2 \left(22 \zeta_2 \zeta_3-39 \zeta_5+\frac{275 \zeta_4}{4}+\frac{1865 \zeta_3}{18} +\frac{3191 \zeta_2}{72}-\frac{98671}{648}\right)\nonumber\\
   &\quad\quad\quad+N_c n_f \left(-\frac{19 \zeta_4}{2}-\frac{157 \zeta_3}{9}-\frac{871
   \zeta_2}{108}+\frac{149033}{3888}\right)\nonumber\\
   &\quad\quad\quad+\frac{n_f}{N_c} \left(3 \zeta_4+\frac{19 \zeta_3}{3}+\frac{\zeta_2}{4}-\frac{1711}{144}\right)+n_f^2 \left(\frac{5 \zeta_2}{54}-\frac{140}{243}\right)\Bigg] \nonumber\\
   &\quad+\epsilon ^2 \Bigg[N_c^2 \left(-\frac{4733 \pi ^6}{30240}-\frac{659 \zeta_3^2}{18}-\frac{8987 \zeta_2 \zeta_3}{36}-\frac{187 \zeta_5}{5}+\frac{16103 \zeta_4}{64}+\frac{121859 \zeta_3}{324}+\frac{71263 \zeta_2}{648}\right.\nonumber\\
   &\quad\quad\quad \left.-\frac{6140957}{11664}\right)+N_c n_f \left(\frac{781 \zeta_2 \zeta_3}{18}+\frac{104 \zeta_5}{5}-\frac{5803 \zeta_4}{144}-\frac{5698 \zeta_3}{81}-\frac{1645 \zeta_2}{72}+\frac{3197809}{23328}\right)\nonumber\\
   &\quad\quad\quad+\frac{n_f}{N_c}
   \left(-2 \zeta_2 \zeta_3+14 \zeta_5+\frac{19 \zeta_4}{2}+\frac{197 \zeta_3}{9}+\frac{55 \zeta_2}{24}-\frac{42727}{864}\right)\nonumber\\
&\quad\quad\quad+n_f^2
   \left(-\frac{5 \zeta_4}{144}+\frac{35 \zeta_3}{81} + \frac{53 \zeta_2}{162}-\frac{404}{243}\right)\Bigg] + \mathcal{O}(\ep^3)\, . \label{eq:impact_gluon_2}
   \end{align}
In order to express the gluon Regge trajectory, we define
\begin{equation}\label{eq:kappa}
K(\as(\mu)) = - \frac{1}{4} \int_\infty^{\mu^2} \frac{d\lambda^2}{\lambda^2} \gamma^\text{K}\left(\as(\lambda^2)\right),
\end{equation}
with the perturbative expansion $K = \sum_{\ell \geq 1} K_\ell \aso^\ell$. The coefficients up to third order are
\begin{align}
    K_1 &= \frac{\gamma_0^\text{K}}{\ep}\, , \nonumber\\
    K_2 &= \frac{2\gamma_1^\text{K}}{\ep} - \frac{\beta_0 \gamma_0^\text{K}}{2 \ep^2}\, ,\nonumber\\
    K_3 &= \frac{16\gamma_2^\text{K}}{3\ep} - \frac{4\beta_0 \gamma_1^\text{K} + 4\beta_1 \gamma_0^\text{K}}{3\ep^2} + \frac{\beta_0^2\gamma_0^\text{K}}{3\ep^3} \, .
\end{align}
The expansion coefficients of the gluon Regge trajectory $\tau_\ell$
can then be written as \cite{Caola:2021izf,Falcioni:2021dgr}
\begin{align}
\label{eq:Regge_gluon_trajectory}
\tau_1 &= \; e^{\ep \gamma_E}  \frac{\Gamma(1-\ep)^2 \Gamma(1+\ep)}
{\Gamma(1-2\ep)} \frac{2}{\ep},  \nonumber\\[8pt]
\tau_2 &= \; K_2  -\frac{56 n_f}{27} 
 + N_c
\left( \frac{404}{27} - 2\zeta_3\right) +\epsilon \bigg[N_c \left(\frac{2428}{81}-66 \zeta_3-\frac{67 \zeta_2}{9}-3\zeta_4\right) \nonumber \\
&\quad+ n_f
  \left(12 \zeta_3 -\frac{328}{81}+\frac{5
  \pi ^2}{27}\right) \bigg]+\epsilon^2 \bigg[N_c \bigg(82\zeta_5
+\frac{142 \zeta_2 \zeta_3}{3}-\frac{4556 \zeta_3 }{27}+\frac{14576}{243}\nonumber \\
&\quad -\frac{404 \zeta_2}{27}-\frac{2321 \zeta_4}{24}\bigg) +n_f \left(\frac{680
  \zeta_3}{27}-\frac{1952}{243}+\frac{56 \zeta_2}{27}+\frac{211 \zeta_4}{12}\right)\bigg] + \mathcal{O}(\ep^3), \nonumber\\[8pt]
  \tau_3 &= \;  K_3 +N_c^2 \bigg(16 \zeta_5 
+\frac{40 \zeta_2 \zeta_3}{3}-\frac{77 \zeta_4}{3}-\frac{6664 \zeta_3}{27} -\frac{3196
  \zeta_2}{81}+\frac{297029}{1458}\bigg) 
+ n_f^2
  \left(\frac{928}{729}-\frac{128 \zeta_3}{27}\right)
\nonumber\\
&\quad+N_c n_f
  \left(\frac{412 \zeta_2}{81}+\frac{2 \zeta_4}{3}+\frac{632 \zeta_3}{9}-\frac{171449}{2916}\right) +\frac{n_f}{N_c} \left(-4
  \zeta_4-\frac{76 \zeta_3}{9}+\frac{1711}{108}\right) +
\mathcal{O}(\ep).
\end{align}
Note that since one can expand $\tau_1 = K_1 + O(\epsilon)$, the poles of $\tau_g$ are given exactly by $K$ defined in eq.~\eqref{eq:kappa} (see also ref.~\cite{Falcioni:2021buo}).

The expressions above are also provided in electronic format in the \texttt{arXiv} submission of this article.

\end{document}